\documentclass[11pt]{article}
\usepackage[top=30truemm,bottom=30truemm,left=30truemm,right=30truemm]{geometry}

\usepackage{graphicx}
\usepackage{multicol,multirow}
\usepackage{amsmath,amssymb,amsfonts}
\usepackage{mathrsfs}
\usepackage{amsthm}
\usepackage{rotating}
\usepackage{appendix}
\usepackage[numbers]{natbib}
\usepackage{amsmath} 
\usepackage{ifpdf}
\usepackage[T1]{fontenc}
\usepackage{newtxtext}
\usepackage{newtxmath}
\usepackage{textcomp}
\usepackage{xcolor}
\usepackage{lipsum}
\usepackage[colorlinks,allcolors=blue]{hyperref}
\usepackage{listings}
\usepackage{subcaption}
\usepackage{caption}
\usepackage{hyperref}
\usepackage{algorithm}
\usepackage{algpseudocode}
\usepackage{float}
\usepackage{placeins}
\usepackage{comment}
\usepackage{bm}

\theoremstyle{definition}

\numberwithin{equation}{section}

\newcommand{\Dim}{D}

\newcommand{\bucketIndex}[1]{b_{#1}}

\newcommand{\VariableSelectionHat}[0]{\hat{S}}
\newcommand{\VariableSelectionHatIndex}[1]{\hat{S}_{#1}}
\newcommand{\TimeN}[0]{T}
\newcommand{\ratioTrain}[0]{\rho_{\rm train}}

\newcommand{\threeDots}[0]{...}
\newcommand{\codeName}[1]{{\tt #1}}

\newcommand{\bx}{{\bm x}}
\newcommand{\by}{{\bm y}}

\title{Variable Selection for Comparing High-dimensional Time-Series Data}

\author{
  Kensuke Mitsuzawa\thanks{Data Science Department, EURECOM, Biot, Alpes Maritimes, France. \texttt{{kensuke.mitsuzawa, motonobu.kanagawa}@eurecom.fr}} \and
  Margherita Grossi\thanks{Intelligent Cloud Technologies Laboratory, Huawei Munich Research Center, München, Bayern, Germany. \texttt{{margherita.grossi, stefano.bortoli}@huawei.com}} \and
  Stefano Bortoli\footnotemark[2] \and
  Motonobu Kanagawa\footnotemark[1]
}

\date{}  

\begin{document}

\maketitle

\begin{abstract}
Given a pair of multivariate time-series data of the same length and dimensions, an approach is proposed to select variables and time intervals where the two series are significantly different. 
In applications where one time series is an output from a computationally expensive simulator, the approach may be used for validating the simulator against real data, for comparing the outputs of two simulators, and for validating a machine learning-based emulator against the simulator. 
With the proposed approach, the entire time interval is split into multiple subintervals, and on each subinterval, the two sample sets are compared to select variables that distinguish their distributions and a two-sample test is performed.
The validity and limitations of the proposed approach are investigated in synthetic data experiments.
Its usefulness is demonstrated in an application with a particle-based fluid simulator, where a deep neural network model is compared against the simulator, and in an application with a microscopic traffic simulator, where the effects of changing the simulator's parameters on traffic flows are analysed. 
\end{abstract}

\section*{Impact Statement}
When engineering experiments on the system of interest are impossible or costly, computer experiments are performed on a simulator as a substitute for the intended experiments.
Such situations can be widely seen in engineering for manufacturing, transportation and urban design, and disaster prevention.  
To trust the results of computer experiments, however, the simulator must be systematically validated beforehand.
Similarly, when a machine learning model is trained to emulate a simulator to accelerate test-time execution, the emulator must be validated against the simulator before the emulator is deployed.
In these cases, a systematic validation is challenging because the simulator's output is usually a high-dimensional time series. 
The proposed approach is expected to be helpful in these situations for performing a systematic validation.

\section{Introduction}

Consider the following problem: {\em Given two multivariate time-series data of the same length and dimensions, identify time intervals and variables where they are significantly different}.
This problem will be referred to as {\em variable selection for two multivariate time series}.
More specifically, given two time series ${\bm x}_1, \dots, {\bm x}_T \in \mathbb{R}^D$ and ${\bm y}_1, \dots, {\bm y}_T \in \mathbb{R}^D$, where $T$ is the number of time points and $D$ is the number of variables, the problem is to identify the subintervals of the entire interval $[1, T] := \{1, \dots, T\}$ and variables $d =  1, \dots, D$ where the two time series are significantly different.

This problem is motivated from the following situations involving {\em simulators} \citep[e.g.,][]{law2007simulation}.
\begin{itemize}
    \item {\bf Model Validation.} One multivariate time series may be output from a simulator of a spatio-temporal phenomenon (e.g., fluid, traffic, etc.), and the other may be real observed data.
    In this case, one can understand {\em when} and {\em where} the simulator fails to model reality by solving the above problem.
    Such insights may be useful in deciding whether the simulator is acceptable for an intended application (e.g., forecasting) and may provide hints on improving the simulator. 

    \item {\bf Model Comparison.} One multivariate time series may be output from a simulator under one scenario (e.g., a traffic simulator during rush hours), and the other may be output from the same simulator under a slightly modified scenario (e.g., some roads are blocked.). 
    Solving the above problem, one can understand when and where this modification influences.

    \item {\bf Emulator Validation.}
    One multivariate time series may be output from a simulator, and the other may be from a machine learning model (e.g., deep neural networks) that emulates the simulator. Solving the above problem enables identifying when and where the emulator fails to imitate the ground-truth simulator.
    
\end{itemize}
In this way, the above problem appears in many scientific and engineering fields where simulators are developed to model spatio-temporal phenomena of interest, such as climate science, fluid mechanics, and digital twin~\citep{sacks-2020, lu-2020}.  
There, the number $D$ of variables can be high  (e.g., hundreds to thousands), as it is usually the number of spatial locations or sensors on which observations are simulated.
Manually analysing such data to detect the differing variables is time-consuming and cannot be systematic. 
Thus, an automatic method for variable selection is desired.

One possible approach might be formulating the problem as a {\em two-sample variable selection} problem (see Sections~\ref{sec:related-work} and \ref{sec:proposal-variable-selection}), but this requires {\em multiple} realisations of each of the two multivariate time series 
$X = ({\bm x}_1, \dots, {\bm x}_T) \in \mathbb{R}^{D \times T}$ and $Y = ({\bm y}_1, \dots, {\bm y}_T) \in \mathbb{R}^{D \times T}$. 
Here, we identify each multivariate time series as a $D \times T$ matrix.
More precisely, if we are given independently and identically distributed (i.i.d.) samples $X_1, \dots, X_n \in \mathbb{R}^{D \times T}$ of $X$ and i.i.d.~samples $Y_1, \dots, Y_m \in \mathbb{R}^{D \times T}$ of $Y$, where $n$ and $m$ are sample sizes, a two-sample variable selection method can select variable-time pairs $(d,t)$ on which the underlying distributions of $X$ and $Y$ differ. 
However, this approach cannot be directly used in the above problem where only a {\em single} pair of multivariate-time series $X$ and $Y$ are given. 

For example, suppose that $X$ is output from a computationally expensive simulator, which takes one day to finish the simulation. 
In this case, simulating $n$ samples $X_1, \dots, X_n$ requires $n$ days, so $n$ cannot be large. 
However, an accurate two-sample testing (and variable selection) would require at least roughly $n = 50$ samples, and this requires 50 days of simulations, which may be unfeasible in practice. 
Such a computationally expensive simulator is common, since generally a simulator's cost increases arbitrarily as  its resolution becomes higher. 
In such a situation, assuming many samples of $X$ and $Y$ is not realistic.

The present work proposes an approach to variable selection for two multivariate time series. 
The proposed approach, named ``Time-Slicing Variable Selection," divides the entire time interval into multiple subintervals, and performs two-sample variable selection on each subinterval to compare the two time series (see Figure~\ref{fig:illust-proposed-method}). 
This is a meta algorithm in the sense that any existing two-sample variable selection method may be combined with the proposed framework. 
As a specific example, the proposed approach using a kernel-based two-sample variable selection method \citep{mitsuzawa-2023} is demonstrated to be effective for the problem, compared with baselines. 
This is demonstrated in synthetic data experiments and two applications.
One application involves fluid simulations, where a simulator and its emulator based on a neural net are compared (emulator validation). 
The other application focuses on traffic simulations, where one scenario and its modified scenario are compared (model comparison). 
Through these experiments, we show that the proposed approach can perform interpretable variable selection only from a single pair of two time series data.

The paper proceeds as follows.
After briefly reviewing related studies in Section~\ref{sec:related-work},
we describe the proposed approach in Section~\ref{sec:proposed-method} and explain two-sample variable selection methods used in our experiments in   Section~\ref{sec:proposal-variable-selection}. 
Section~\ref{sec:experiment} reports synthetic data experiments, Section~\ref{sec:demonstration} on fluid simulations, and Section~\ref{sec:demonstration-most} on traffic simulations.

\subsection{Related Work}
\label{sec:related-work} 

To our knowledge, the problem of variable selection for two multivariate time series is new. 
Thus, we review existing papers that study problems related to (but not exactly the same as) our problem. \\

\noindent
{\bf Comparison of Multivariate Time Series.} 
\label{sec:related-comp-multi-var}
\citet{tapinos2013method} propose a semi-metric to compare two multivariate time series with different dimensions.  
To compare two simulation outputs, 
\citet{cess2022representation} propose to train a neural network for each output to project it into a shared lower-dimensional space, and compute the distance between the two projections. 
\citet{campbell2006sensitivity} study sensitivity analysis of a simulator whose output is a function of, e.g., time and space.  Each output function is expanded using finite basis functions, and sensitivity analysis is performed on the functions' coefficients.
\citet{whynne-2022} propose a two-sample test for testing the null hypothesis that two functional data are generated from the same process.
These works assume multiple samples for each time series are available and do not study variable selection.
Thus, their settings are different from ours.  
\\

\noindent
{\bf Two-Sample Variable Selection.}
\label{sec:variable-selection-for-two-sample-testing}
The task of two-sample variable selection may be defined as follows.
Let $P$ and $Q$ be probability distributions on the $D$-dimensional Euclidean space $\mathbb{R}^D$.
Suppose i.i.d.~samples ${\bm x}_1, \dots, {\bm x}_n \in \mathbb{R}^D$ from $P$ and ${\bm y}_1, \dots, {\bm y}_m \in \mathbb{R}^D$ from $Q$ are provided.  
Then the task is to select a subset of variables $S \subset \{ 1, 2, \dots, D \}$ responsible for any discrepancies between $P$ and $Q$ (if they are different).
For a more rigorous definition, see \citet[Section 3.1]{mitsuzawa-2023}.
Several methods exist for two-sample variable selection \citep{hido-2008,lopez-paz2017,sutherland2017, mitsuzawa-2023,yamada-2018,Lim-2020,mueller-2015}.
Any of them can be combined with the proposed framework. Section~\ref{sec:proposal-variable-selection} describes those methods studied in the experiments. 
For a systematic review, we refer to \citet[Section 2]{mitsuzawa-2023}.

\section{Proposed Framework}
\label{sec:proposed-method}

  \begin{figure}[t]
      \centering
      \includegraphics[width=0.9\linewidth]{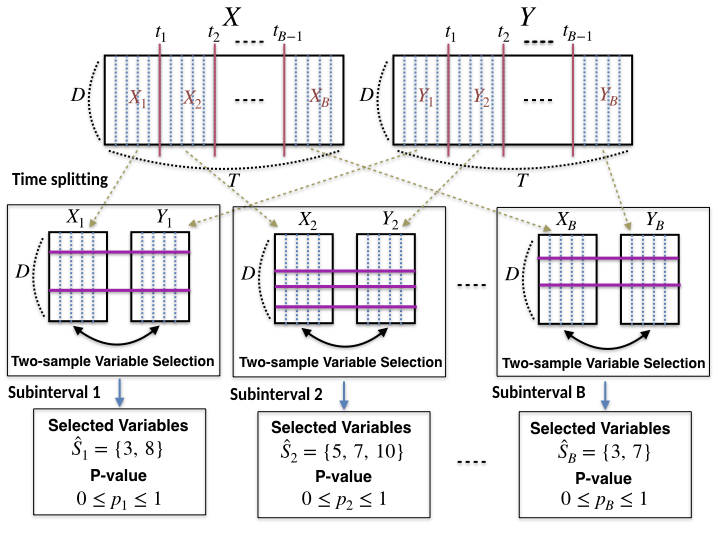}
      \caption{Illustration of the proposed approach to variable selection for two multivariate time series. 
      {\bf Top row:} $X = (\bx_1, \dots, \bx_T) \in \mathbb{R}^{D \times T}$ and $Y = (\by_1, \dots, \by_T) \in \mathbb{R}^{D \times T}$ represent two given multivariate time series data. 
      Time-splitting is applied to each of $X$ and $Y$ at time points $t_1, t_2, \dots, t_{B-1}$.
      {\bf Middle row:} On each subinterval $b = 1, 2, \dots, B$, two-sample variable selection is applied to compare the data matrices $X_b = (\bx_{t_{b-1}+1}, \dots, \bx_{ t_b })$ and $Y_b = (\by_{t_{b-1}+1}, \dots, \by_{ t_b })$.
      {\bf Bottom row:} Selected variables $\hat{S}_b \subset \{1, \dots, D\}$ from the middle row are used to perform a permutation two-sample test to calculate a p-value $p_b$.
      }
      \label{fig:illust-proposed-method}
  \end{figure}

  This section describes the proposed approach to variable selection for comparing two multivariate time series. 
  We define the setting in Section~\ref{sec:data-and-purpose}, explain our approach in Section~\ref{sec:time-slicing-variable-selection}, and discuss it with examples in Section~\ref{sec:example-time-slicing-variable-selection}.

  \subsection{Data and Problem}
  \label{sec:data-and-purpose}
  
  Suppose we have two $D$-by-$T$ matrices $X \in \mathbb{R}^{D \times T}$ and $Y \in \mathbb{R}^{D \times T}$, each representing a time-series of $D$-dimensional vectors of length $T$:
  \begin{align*}
  X &= (\bx_1, \bx_2, \dots, \bx_T) \in \mathbb{R}^{D \times T}, \quad \text{where}\quad \bx_t = (x_{t,1}, \dots, x_{t,D})^\top \in \mathbb{R}^D \quad (t = 1, \dots, T), \\
  Y &= (\by_1, \by_2, \dots, \by_T)  \in \mathbb{R}^{D \times T}, \quad \text{where}\quad \by_t = (y_{t,1}, \dots, y_{t,D})^\top \in \mathbb{R}^D \quad (t = 1, \dots, T).
  \end{align*}
  For example, $X$ may be generated from a traffic simulator, and $Y$ from the same simulator but under a different parameter setting.
  Each element, $x_{d,t}$ or $y_{d,t}$, is a summary statistic of observations, such as traffic flow, from the $d$-th sensor (where $d = 1, \dots, D$) during the $t$-th time interval (where $t = 1,\dots, T$) in the simulated traffic system.
  
  The problem is to identify {\em subintervals} of the entire time interval $[1, T] := (1, 2, \dots, T)$ and {\em variables} $d = 1, \dots, D$ where the two time series $X$ and $Y$ differ significantly. 
  In the traffic simulation example, such time intervals and variables are where the traffic flows differ due to the change of the parameter setting. 
  We will propose a framework for identifying such differing times and variables.

  \subsection{Procedure}
  \label{sec:time-slicing-variable-selection}

  \begin{algorithm}[t]
    \caption{
      Proposed Framework 
      \newline
      \Comment{
        {\bf Input:} Two time-series data $X := \bx_1,\threeDots, \bx_{\TimeN}) \in \mathbb{R}^{D \times T}$ and $Y:= (\by_1,\threeDots,\by_{\TimeN})  \in \mathbb{R}^{\Dim \times \TimeN}$.
        Time-splitting points $0 =: t_0 < t_1 < \cdots <  t_{B-1} < t_B := T$. 
        Ratio $0 < \ratioTrain < 1$ of training-test splitting on each subinterval. 
      }
      \newline
      \Comment{
        {\bf Output:} 
        Selected variables $\hat{S}_b \subset \{1, 2, \dots, D\}$ for each $b = 1,\dots, B$. 
        P-value $p_b \in [0,1]$ of a two-sample test for each $b = 1, \dots, B$.
      }
    }
    \begin{algorithmic}[1]
    \ForAll{$b = 1, \dots, B$}
      \State Set $X_b = (\bx_{t_{b-1} + 1}, \dots, \bx_{t_b}) \in \mathbb{R}^{D \times (t_b - t_{b-1})}$ and $Y_b = (\by_{t_{b-1} + 1}, \dots, \by_{t_b})  \in \mathbb{R}^{D \times (t_b - t_{b-1})}$.
      \State Randomly split $X_b$ into $X_b^{\rm (tr)} \in \mathbb{R}^{D \times (t_b - t_{b-1}) \rho_{\rm train}}$ and  $X_b^{\rm (te)} \in \mathbb{R}^{D \times (t_b - t_{b-1}) (1 - \rho_{\rm train})}$.
        \State Randomly split $Y_b$ into $Y_b^{\rm (tr)} \in \mathbb{R}^{D \times (t_b - t_{b-1}) \rho_{\rm train}}$ and  $Y_b^{\rm (te)} \in \mathbb{R}^{D \times (t_b - t_{b-1}) (1 - \rho_{\rm train})}$.  
      
      \State Perform two-sample variable selection on $X_b^{\rm (tr)}$ and $Y_b^{\rm (tr)}$ to obtain $\hat{S}_b \subset \{1, 2, \dots, D\}$.
      \State Perform a permutation two-sample test on $X_b^{\rm (te)}$ and $Y_b^{\rm (te)}$ using the selected variables $\hat{S}_b$ to obtain a p-value $p_b \in [0,1]$  for the null hypothesis that the marginal distributions of $X_b^{\rm (te)}$ and $Y_b^{\rm (te)}$ on are the same. 
    \EndFor
    \end{algorithmic}
    \label{alg:proposed-framework}
    \end{algorithm}

  We describe here the proposed procedure, which is illustrated in Figure~\ref{fig:illust-proposed-method} and summarised in Algorithm~\ref{alg:proposed-framework}. \\
  
  \noindent
  \underline{\bf 1. Time splitting.}
  The main idea is to split the entire time interval $[1, T]$ into $B$ disjoint subintervals for some $B < T$: 
  \begin{align}
  & [0, T] = [t_0 + 1, t_1] \cup [t_1 + 1, t_2] \cup \cdots \cup [t_{B-1} + 1, t_B],  \label{eq:split-sub-intervals} \\
  & \text{where} \quad 0 =:t_0 < t_1 < t_2 < \cdots < t_{B-1} < t_B := T. \nonumber
  \end{align}
  These are the time points where we split the entire time interval $[1,T]$. 
  Note that here we use the notation $[t_{b-1}+1, t_b] := (t_{b-1} + 1,  t_{b-1} + 2, \dots, t_b)$ for $b = 1, \dots, B$.
  For example, one may split $[1,T]$ into equal-size intervals: If $T = B m $ for some $m \in \mathbb{N}$, one sets 
  $$
  t_1 = m,\quad t_2 = 2m,\quad \dots,\quad t_{B-1} = (B-1) m,
  $$
  so that each interval $[t_{b-1}+1, t_b]$ consists of $m$ time points.

  \label{sec:step1-time-slicing}
  
  Both time series $X$ and $Y$ are split according to the subintervals~\eqref{eq:split-sub-intervals}. 
  That is, $X$ is split into $B$ sub time-series $X_1,  \dots, X_B$ and $Y$ into $B$ sub time-series $Y_1,  \dots, Y_B$:
  \begin{align*}
      X & = (X_1,  \dots, X_B),\ \    \text{where} \ \  X_b := (\bx_{t_{b-1} + 1}, \dots, \bx_{t_b}) \in \mathbb{R}^{D \times (t_b - t_{b-1})} \quad (b = 1, \dots ,B),   \\
      Y & = (Y_1, \dots, Y_B),\ \  \text{where} \ \  Y_b := (\by_{t_{b-1} + 1}, \dots, \by_{t_b}) \in \mathbb{R}^{D \times (t_b - t_{b-1})} \quad (b = 1, \dots, B).   
  \end{align*}
  For the $b$-th subinterval ($b=1,\dots,B$), the sub time-series $X_b$ is a set of $D$-dimensional  vectors $\bx_{t_{b-1} + 1}, \dots, \bx_{t_b} \in \mathbb{R}^D$, and the sub time-series $Y_b$ is a set of $D$-dimensional  vectors $\by_{t_{b-1} + 1}, \dots, \by_{t_b} \in \mathbb{R}^D$. The time order within each of $X_b$ and $Y_b$ shall be neglected. 
  \\
  
  \noindent
  \underline{\bf 2. Two-sample variable selection.}
  The following is performed for
   each subinterval $b = 1, \dots, B$. 
  Let $\rho_{\rm train} \in (0,1)$ be a constant. 
  Each of $X_b = (\bx_{t_{b-1} + 1}, \dots, \bx_{t_b})$ and $Y_b = (\by_{t_{b-1} + 1}, \dots, \by_{t_b})$ is randomly split into ``training'' and ``test'' subsets in the ratio $\rho_{\rm train}: 1 - \rho_{\rm train}$; denote the resulting subsets by 
  \begin{align*}
  & X_b^{(\rm tr)} \in \mathbb{R}^{D \times (t_b - t_{b-1})\rho_{\rm train}}, \quad X_b^{(\rm te)} \in \mathbb{R}^{D \times (t_b - t_{b-1})(1-\rho_{\rm train})},  \\
  & Y_b^{(\rm tr)} \in \mathbb{R}^{D \times (t_b - t_{b-1})\rho_{\rm train}}, \quad Y_b^{(\rm te)} \in \mathbb{R}^{D \times (t_b - t_{b-1})(1-\rho_{\rm train})}.
  \end{align*}
  Here, we assume $(t_b - t_{b-1})\rho_{\rm train}$ and $(t_b - t_{b-1})(1-\rho_{\rm train})$ are integers for simplicity. 
  
  Two-sample variable selection is then performed on the training sets $X_b^{(\rm tr)}$ and $Y_b^{(\rm tr)}$, selecting a subset of variables on which the distributions of $X_b^{(\rm tr)}$ and $Y_b^{(\rm tr)}$ differ significantly: 
  $$
  \hat{S}_b \subset \{1, 2, \dots, D\}.
  $$ 
  The concrete algorithm for two-sample variable selection is to be chosen by the user of the proposed framework. 
  Section~\ref{sec:proposal-mmd-based-variable-selection} reviews algorithms used in the experiments.
  
  Lastly, for each $b = 1, \dots, B$,  a two-sample permutation test \citep[e.g.,][Chapter 15]{efron1994introduction} is performed on the test sets $X_b^{(\rm te)}$ and $Y_b^{(\rm te)}$ using only the selected variables $\hat{S}_b$, to yield a p-value, 
  $$
  0 \leq p_b \leq 1.
  $$ 
  The null hypothesis is that the marginal distributions of $X_b^{(\rm te)}$ and $Y_b^{(\rm te)}$ on $\hat{S}_b$ are the same. 
  If they are largely different, the p-value becomes small, suggesting that the selected variables distinguish the two time series on this subinterval.
  If the marginal distributions are similar, the p-value becomes large, suggesting that there is not much difference between the two time series on the selected variables on this subinterval.
  In this sense, the p-value informs whether the selected variables distinguish the two time series on this subinterval. 
  
  The concrete test statistic here is to be chosen by the user of the framework (as for two-sample variable selection). 
  The sliced Wasserstein distance \citep{bonneel-2015} is used in our experiments.

  \subsection{Example and Discussion}
  \label{sec:example-time-slicing-variable-selection}

  \begin{figure}[t]
      \centering
      \includegraphics[scale=0.3]{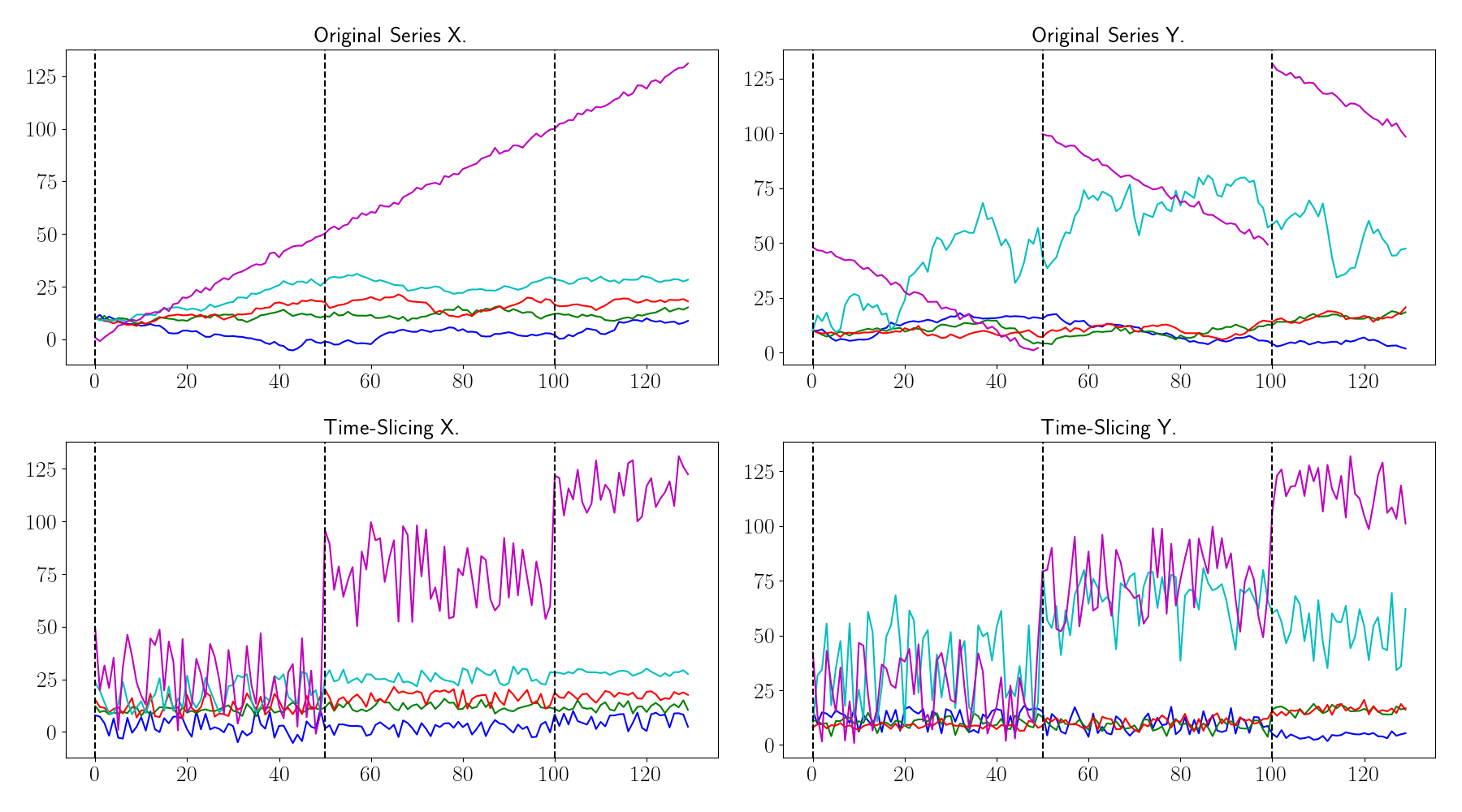}
      \caption{
      An example of time-splitting applied to two time-series data $X = (\bx_1, \dots, \bx_T) \in \mathbb{R}^{D \times T}$ and $Y = (\by_1, \dots, \by_T) \in \mathbb{R}^{D \times T}$ with $D = 5$ and $T = 130$, with $t_1 = 50$, $t_2 = 100$ and $t_3 = 130 = T$. 
      The top plots show the two-time series $X$ and $Y$, where 5 different colours correspond to the 5 variables (or dimensions).
      In each plot, the horizontal axis represents time points, and the vertical axis the values of each variable.
      The two variables represented by the violet and light-blue colours follow different stochastic processes for $X$ and $Y$. 
      The bottom figures show the results of applying the time-splitting and randomisation in each subinterval. 
      }
      \label{fig:example-time-slicing-iid-idea-b}
  \end{figure}

  Figure~\ref{fig:example-time-slicing-iid-idea-b} shows an example of the time-splitting operation. 
  The top tow figures describe two given time-series data $X = (\bx_1, \dots, \bx_T) \in \mathbb{R}^{D \times T}$ and $Y = (\by_1, \dots, \by_T) \in \mathbb{R}^{D \times T}$ with $D = 5$ variables and $T = 130$ time points. 
  The two variables indicated by the violet and light-blue trajectories are generated from different stochastic processes for $X$ and $Y$. 
  
  We split each time-series into $B = 3$ subintervals, with $t_1 = 50$, $t_2 = 100$ and $t_3 = 130$, which are shown in the bottom two figures. 
  Here, on each subinterval $b = 1, \dots, B$, we randomise the time order of data vectors within $X_b = (\bx_{t_{b-1} + 1}, \dots, \bx_{t_b})$ and those within $Y_b = (\by_{t_{b-1} + 1}, \dots, \by_{t_b})$.
  This is to illustrate that, by applying two-sample variable selection and two-sample testing on $X_b$ and $Y_b$, {\em we essentially treat data vectors in $X_b$ (and those in $Y_b$) as independent}, since we break the time structure within $X_b$ and that within $Y_b$ when applying these approaches (see Section~\ref{sec:synthetic-exp} for details).

  Let us discuss how the proposed approach would work for this example. 
  \begin{itemize}
      \item Regarding the light-blue trajectories, whose underlying stochastic processes are different for $X$ and $Y$, the empirical distributions of $X_b$ and $Y_b$ for this variable are noticeably different for each subinterval $b = 1, 2, 3$. 
  Therefore, two-sample variable selection would be able to identify this variable as differing for each of $b = 1, 2, 3$. 
  
  \item The violet trajectories, which clearly differ for $X$ and $Y$, are deliberately constructed to illustrate an ``unfortunate" situation where time-slicing makes the two time-series indistinguishable.
  After the time-slicing, the empirical distributions of $X_b$ and $Y_b$ for this variable become very similar (actually, here, we constructed them so that they become exactly the same) for each interval $b = 1, 2, 3$.  
  In this case, the proposed approach would not be able to detect the difference between the two time-series for this variable. 
  
  \end{itemize}

  The issue with the violet trajectories could have been avoided if there were many more time-splicing points (i.e., $B$ is larger) so that each subinterval was shorter and thus could capture the difference in the two trajectories.
  This is demonstrated in the experiments in Section~\ref{sec:synthetic-exp}.

  This example suggests that the proposed framework would work well if 
  \begin{enumerate}
      \item the time length of each subinterval $[t_{b-1}+1, t_b]$ is short enough so that the $X_b = ( \bx_{t_{b-1}+1}, \dots, \bx_{t_{b}} )$ and $Y_b = ( \by_{t_{b-1}+1}, \dots, \by_{t_{b}} )$ are approximately stationary, while 
      \item the number of sample vectors $t_b - t_{b-1}$ in $X_b$ and $Y_b$ is large enough so that two-sample variable selection can be performed reliably.  
  \end{enumerate}
  In point 1, the ``time length'' means that in the physics sense; recall that $t_{b} - t_{b-1}$ is the {\em number} of time steps within the subinterval, not the time length.
  For example, suppose we consider a traffic simulation from 8 am to 9 am and set the total number of time steps to $T = 1,000$. 
  Then the total time length is $3,600$ seconds, and the time length of one time step is $3,600 / 1,000 = 3.6$ seconds. 
  If the number of subintervals is $B = 4$, and the slicing points are $t_1 = 250$, $t_2 = 500$, $t_3 = 750$ and $t_4 = 1,000$, then the time length of each subinterval is $3,600 / 4 = 900$ seconds, and the number of time steps within the subinterval is $250$.
  
  There is a trade-off between points 1 and 2 above. 
  If we make the time length of each interval shorter (or longer), then the time steps within one interval becomes smaller (or larger).  
  For example, in the above example, if we change the number of subintervals to $B = 40$, then the time length of each subinterval is $90$ seconds, while the number of time steps within the subinterval is $25$. 
  In this case, the traffic flows within each subinterval would remain unchanged (and thus can be regarded as stationary), while we only have a small number of observations, making statistical estimation more challenging.

  \label{sec:proposed-method-limitation}

  \section{Two-Sample Variable Selection Algorithms}
  \label{sec:proposal-variable-selection}

  We briefly describe two-sample variable selection algorithms that we try within the proposed framework in our experiments; for details, we refer to the corresponding references.
  Except for one method, each algorithm computes $D$-dimensional weights that represent the importance of individual variables $d = 1, \dots, D$. 
  From these weights, we obtain selected variables $\hat{S} \subset \{1, \dots, D\}$ by using the histogram-based thresholding algorithm of 
  ~\citet[Section 4.2.1]{mitsuzawa-2023}.

  \subsection{MMD-based Two-Sample Variable Selection}
  \label{sec:proposal-mmd-based-variable-selection}

  Maximum Mean Discrepancy (MMD) is a kernel-based distance metric between probability distributions~\citep{gretton-2012a}. 
  MMD offers a way of measuring the distance between two datasets, and thus has been used for model comparison and validation of simulations; see e.g.~\citet{sanchez-2020,feng2023trafficgen} for applications to traffic and particle simulations.
  Recently, two-sample variable algorithms based on MMD have been studied in the literature 
  \citep{sutherland2017,wang2023variable,mitsuzawa-2023}.
  Here, we explain the approach of \cite{mitsuzawa-2023}.
  
  For a kernel function $k: \mathbb{R}^D \times \mathbb{R}^D \to \mathbb{R}$, the MMD between probability distributions $P$ and $Q$ on $\mathbb{R}^D$ is defined as   
  \begin{equation*}
      {\rm MMD}_k^2(P, Q) := \mathbb{E}_{\bx, \bx' {\sim} P}[k(\bx, \bx')] + \mathbb{E}_{\by, \by' \sim Q}[k(\by, \by')]  - 2\mathbb{E}_{\bx \sim P,\ \by \sim Q}[k(\bx, \by)],  
  \end{equation*}
  where $\bx, \bx' \stackrel{i.i.d.}{\sim} P$ and  $\by, \by' \stackrel{i.i.d.}{\sim} Q$ are random vectors, and the expectation $\mathbb{E}$ is taken for the random vectors in the subscript. 
  This is a proper distance metric on probability distributions, if the kernel $k$ is characteristic (e.g., when $k$ is Gaussian) \citep{SriGreFukSchetal10}. 
  Given i.i.d.~samples ${\bf X} = \{ \bx_1, \dots, \bx_n \} \stackrel{i.i.d.}{\sim} P$ and ${\bf Y} = \{\by_1, \dots, \by_n\} \stackrel{i.i.d.}{\sim} Q$, the MMD can be consistently estimated as  
  \begin{equation} 
    \label{eq:mmd-unbiased-est}
    \begin{split}
      & \widehat{\rm MMD}^{2}_n( {\bf X}, {\bf Y}) := \frac{1}{n (n-1)} \sum_{i \not= i'} k(\bx_i, \bx_{i'})  + \frac{1}{n(n-1)} \sum_{ j \not= j'} k(\by_j, \by_{j'})  - \frac{2}{n^2} \sum_{i,j} k(\bx_i, \by_j).
    \end{split}
  \end{equation}
  
  For two-sample variable selection, one can use the {\em Automatic Relevance Detection (ARD)} kernel as the kernel $k$ in \eqref{eq:mmd-unbiased-est}, which is defined as 
  \begin{equation} \label{eq:ARD-kernel}
          k(\bx, \by) = \exp \left(- \frac{1}{D} \sum_{d=1}^{D} \frac{a_d^2 (x_d - y_d)^2}{\gamma^2_d} \right), \quad \bx:=(x_1,\threeDots,x_D)^\intercal \in \mathbb{R}^{D},\ \by:=(y_1,\threeDots,y_D)^\intercal \in \mathbb{R}^{D}
  \end{equation}
  where  $a_1,\threeDots,a_D \geq 0$ are called {\em ARD weights}, and $\gamma_1, \dots, \gamma_D > 0$ are constants to unit-normalise each variable. 
  We compute $\gamma_1, \dots, \gamma_D$ by using the {\em dimension-wise median (or mean) heuristic}; see \citet[Appendix C]{mitsuzawa-2023} for details.

  \citet{sutherland2017} proposed to
  optimise the ARD weights $a_1, \dots, a_D$ in \eqref{eq:ARD-kernel} so as to maximise the {\em test power} (i.e., the rejection probability of the null hypothesis $P = Q$ when $P \not=Q$) of a two-sample test based on the MMD estimate~\eqref{eq:mmd-unbiased-est}.
  \citet{mitsuzawa-2023} extended this approach by introducing $L_1$ regularisation \citep{Tibshirani-1996} to encourage sparsity for variable selection: 
  \begin{equation}
    \label{eq:mmd-optimisation-problem}
    \underset{a \in \mathbb{R}^{\Dim}}{\min} -\log(\ell(a_1, \threeDots, a_{\Dim})) + \lambda \sum_{d=1}^{\Dim} |a_d|, \quad \text{where}\quad \ell(a_1, \threeDots, a_{\Dim}) := \frac{ \widehat{\rm MMD}^{2}_n( {\bf X}, {\bf Y})}{ \sqrt{\mathbb{V}_n({\bf X}, {\bf Y}) + C}}
  \end{equation}
  where $\lambda > 0$ is a regularisation constant, $C = 10^{-8}$ is a constant for numerical stability, and $\mathbb{V}_n({\bf X}, {\bf Y})$ is an empirical estimate of the variance of $\widehat{\rm MMD}^{2}_n( {\bf X}, {\bf Y})$; see \citet{mitsuzawa-2023} for the concrete form. 
  
  The weights $a_1, \dots, a_D$ with larger $\ell(a_1, \dots, a_D)$ makes the test power higher, thus making the two data sets ${\bf X}$ and ${\bf Y}$ more distinguishable. 
  Minimizing the negative logarithm of $\ell(a_1, \dots, a_D)$ plus the $L_1$ regularisation term $\lambda \sum_{d=1}^{\Dim} |a_d|$ thus yield {\em sparse} weights $a_1, \dots, a_D$ that distinguish ${\bf X}$ and ${\bf Y}$. 
  Such weights can be used for selecting variables where the two data sets differ significantly. 
  
  The regularisation constant $\lambda$ controls the sparsity of optimised ARD weights. 
  Thus, an appropriate value of $\lambda$ is unknown beforehand, as it depends on the number of unknown active variables $S \subset \{1, \dots, D \}$ on which $P$ and $Q$ are different. 
  \citet{mitsuzawa-2023} proposed two methods for addressing the regularisation parameter selection.
  One method, referred to as `MMD Model Selection' ({\tt MMD-Selection}), selects the regularisation parameter in a data-driven way.
  The other method,  called  `MMD Cross-Validation Aggregation' ({\tt MMD-CV-AGG}), aggregates the results of using different regularization parameters. 
  For details, we refer to \cite{mitsuzawa-2023}.  
  Implementation details for our experiments are available in Appendix~\ref{sec:app-MMD}.

  \subsection{Variable Selection by Marginal Distribution Comparisons}
  
  One approach to two-sample variable selection is based on the comparisons of one-dimensional marginal distributions.  
  To describe this, let $P$ and $Q$ be probability distributions on $\mathbb{R}^D$.
  For $d = 1, \dots, D$, let $P_d$ and $Q_d$ be the marginal distributions of $P$ and $Q$ on the $d$-th variable. 
  Let ${\rm dist}(\mu, \nu)$ be a distance metric between two probability distributions $\mu$ and $\nu$ on $\mathbb{R}$, such as the MMD and the Wasserstein distance~\citep[e.g.,][]{villani2009optimal}.
  We then define a weight $w_d$ for each variable $d = 1,\dots, D$ as  the distance of each pair of one-dimensional marginal distributions $P_d$ and $Q_d$:
  \begin{equation} \label{eq:marginal-weights-499}
  w_d := {\rm dist}(P_d, Q_d) \quad (d = 1,\dots, D).
  \end{equation}

  In practice, one needs to estimate these weights using samples $\bx_1, \dots, \bx_n \stackrel{i.i.d.}{\sim} P$ and $\by_1, \dots, \by_n \stackrel{i.i.d.}{\sim} Q$, where $\bx_i = ( x_{i,1}, \dots, x_{i,D} )^\top \in \mathbb{R}^D$ and $\by_i = ( y_{i,1}, \dots, y_{i,D} )^\top \in~\mathbb{R}^D$ for $i = 1, \dots, n$. 
  For each $d = 1, \dots, D$, let $\hat{P}_d := \frac{1}{n} \sum_{i=1}^n \delta_{ x_{d,i} }$ be the empirical distribution of $x_{1,d}, \dots, x_{n,d} \in \mathbb{R}$, and $\hat{Q}_d := \frac{1}{n} \sum_{i=1}^n \delta_{ y_{i,d} }$ be the empirical distribution of $y_{1,d}, \dots, y_{n,d} \in \mathbb{R}$, where $\delta_z$ for $z \in \mathbb{R}$ denotes the Dirac distribution at $z$. 
  The $\hat{P}_d$ and $\hat{Q}_d$ are respectively consistent approximations of $P_d$ and $Q_d$. 
  Thus, one can estimate the weights~\eqref{eq:marginal-weights-499} by using $\hat{P}_d$ and $\hat{Q}_d$ as
  $$
  \hat{w}_d := {\rm dist}(\hat{P}_d, \hat{Q}_d) \quad (d = 1, \dots, D).
  $$
  We use the following distance metrics in our experiments.
  \begin{enumerate} 
      \item {\bf Wasserstein distance.}  We compute the Wasserstein-$1$ distance between $\hat{P}_d$ and $\hat{Q}_d$, which is identical to the $L_1$ distance between the cumulative distribution functions $\hat{F}_d$ of $\hat{P}_d$ and $\hat{G}_d$ of $\hat{Q}_d$:
      $$
  \hat{w}_d := W_1( \hat{P}_d, \hat{Q}_d ) = \int_{\mathbb{R}} \left| \hat{F}_d(t) - \hat{G}_d (t) \right| dt \quad (d = 1, \dots, D).
  $$
  We compute it using the the implementation of {\tt SciPy} library~\citep{2020SciPy-NMeth}.
  We use the histogram-based thresholding algorithm of \citet[Section 4.2.1]{mitsuzawa-2023} to the weights $\hat{w}_1, \dots, \hat{w}_D$ to select variables $\hat{S} \subset \{1, \dots, D\}$.
  
  \item {\bf Maximum Mean Discrepancy.}
  We consider the approach of \citet{Lim-2020}.
  For each $d = 1,\dots, D$, it computes the MMD between $\hat{P}_d$ and $\hat{Q}_d$,
  $$
  \hat{w}_d := {\rm MMD}_{k_1}(\hat{P}_d, \hat{Q}_d),
  $$
  where the kernel $k_1$ on $\mathbb{R}$ is the Gaussian kernel whose bandwidth is determined by the median heuristic~\citep{damien-2018} or the inverse multi-quadratic (IMQ) kernel.\footnote{More precisely, \citet{Lim-2020} uses the linear-time MMD estimator of \citet[Section 6]{gretton-2012a}.} 
  From $\hat{w}_1, \dots, \hat{w}_D$, it then selects variables $\hat{S} \subset \{ 1, \dots, D \}$ based on post-selection inference. 
  This approach requires the number of candidate variables to be selected as a hyperparameter. 
  To avoid confusion with the MMD-based approach in Section~\ref{sec:proposal-mmd-based-variable-selection}, we call this method {\tt Mskernel}, following the naming of the authors' code, which we use in our experiments.\footnote{\url{https://github.com/jenninglim/multiscale-features} 
  }

  \end{enumerate}
   
  \label{sec:proposal-wasserstein-based-variable-selection}
  
  \label{sec:proposal-mskernel-variable-selection}

\section{Synthetic Data Experiments}
\label{sec:experiment}

\label{sec:synthetic-exp}
We investigate how the proposed framework in Algorithm~\ref{alg:proposed-framework} works based on synthetically generated data, for which we know ground-truth variables and time points where changes occur.

\subsection{Data Setting}
\label{sec:experiment-triangle}

We set $D = 5$ and $T = 1,000$.
We consider the following two different settings for data generation.\\

\begin{figure}[t]    
    \centering
    \begin{subfigure}[b]{0.3\textwidth}
        \centering
        \raisebox{0.8cm}{\includegraphics[width=\textwidth]{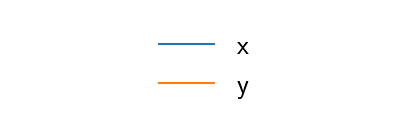}}
    \end{subfigure}
    \hfill
    \begin{subfigure}[b]{0.3\textwidth}
        \centering
        \includegraphics[width=\textwidth]{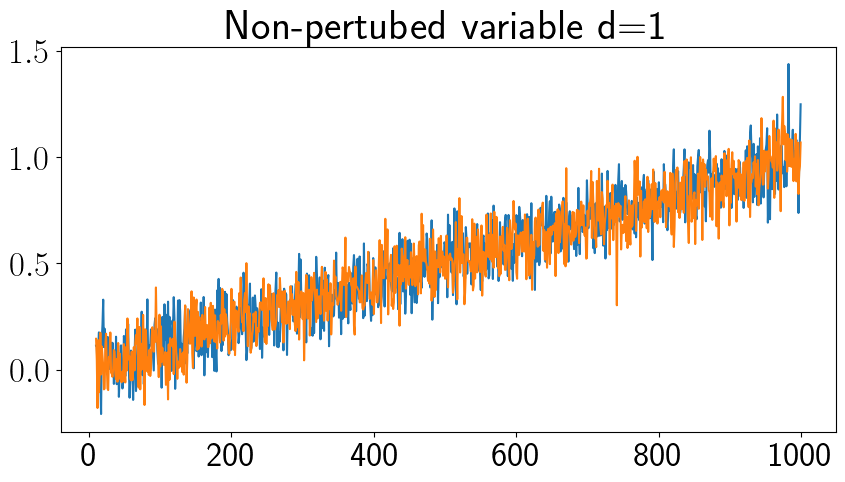}
    \end{subfigure}
    \hfill
    \begin{subfigure}[b]{0.3\textwidth}
        \centering
        \includegraphics[width=\textwidth]{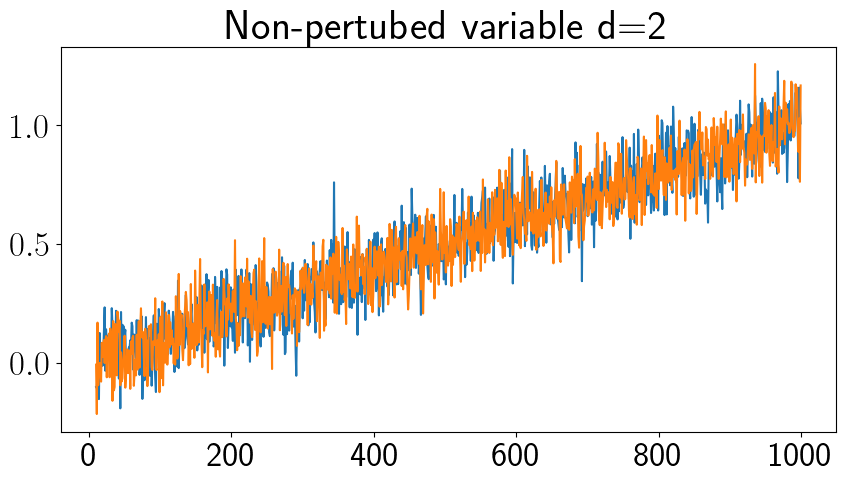}
    \end{subfigure}
    
    \vskip\baselineskip
    
    \begin{subfigure}[b]{0.3\textwidth}
        \centering
        \includegraphics[width=\textwidth]{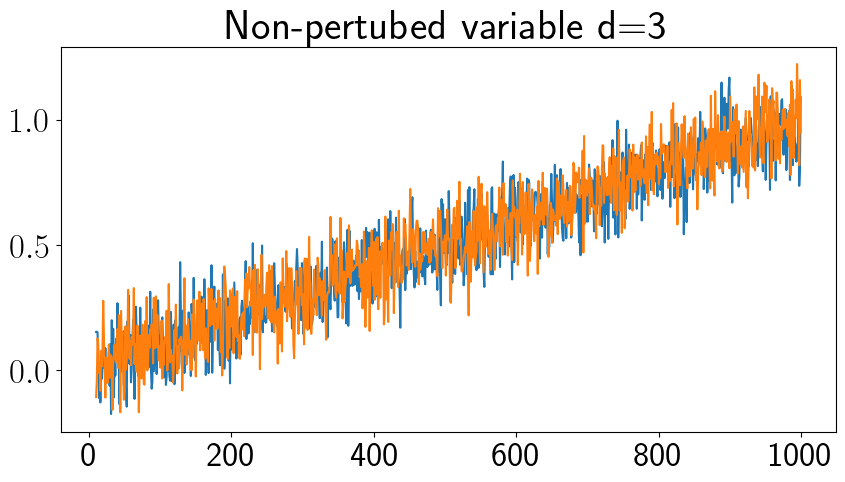}
    \end{subfigure}
    \hfill
    \begin{subfigure}[b]{0.3\textwidth}
        \centering
        \includegraphics[width=\textwidth]{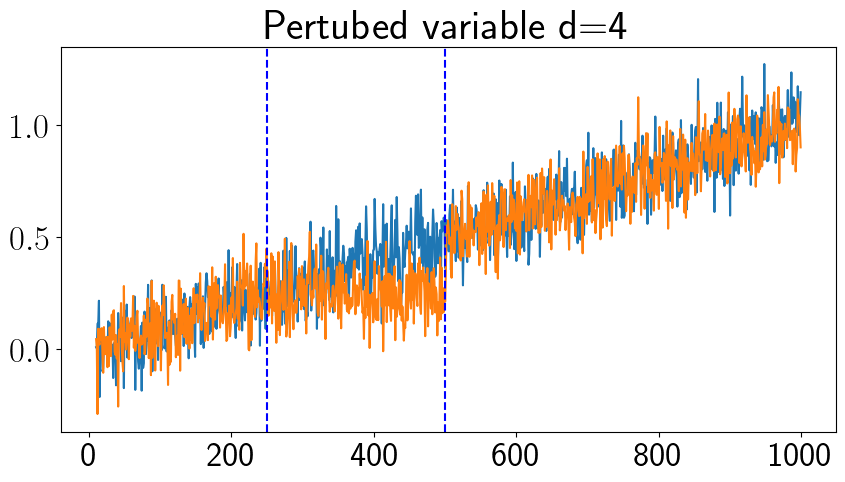}
    \end{subfigure}
    \hfill
    \begin{subfigure}[b]{0.3\textwidth}
        \centering
        \includegraphics[width=\textwidth]{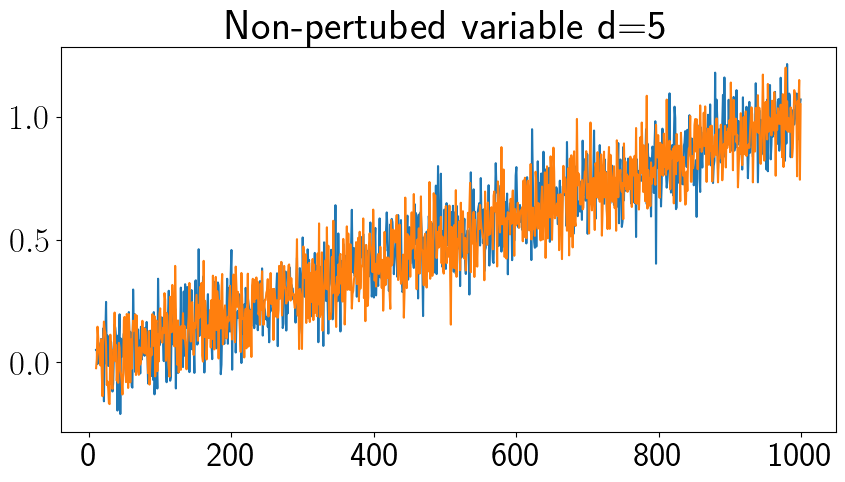}
    \end{subfigure}

    \caption{
    Illustration of two time-series data $X \in \mathbb{R}^{D \times T}$ and $Y  \in \mathbb{R}^{D \times T}$ in  Eq.~\eqref{eq:data-setting-1-56}. 
    Each subfigure shows the trajectories of $X$ and $Y$ in each variable $d= 1,\dots,D$, i.e.,  $x_{1,d}, \dots, x_{T,d}$ and $y_{1,d}, \dots, y_{T,d}$. 
   The variable $d=4$ from $t=251$ to $t = 500$ is where $X$ and $Y$ differ.
    }
    \label{fig:production-2024-03-12-data-visualization}
\end{figure}

\noindent
{\bf Setting 1.} 
We generate two time-series data $X:= (\bx_1, \dots, \bx_{\TimeN}) \in \mathbb{R}^{\Dim \times \TimeN}$ and $Y:= (\by_1, \dots, \by_{\TimeN}) \in \mathbb{R}^{\Dim \times \TimeN}$, where $\bx_t = (x_{t,1}, \dots, x_{t,D})^\top \in \mathbb{R}^D$ and $\by_t = (y_{t,1}, \dots, y_{t,D})^\top \in \mathbb{R}^D$ for $t = 1,\dots, T$, as
\begin{align}
x_{t,d} &= t/T + \epsilon_{t,d}   ~~ \text{for} ~~  d = 1,\dots, D ~~ \text{and}~~  t = 1, \dots, T,  \nonumber  \\
y_{t,d} & = 
\begin{cases}
0.25 + \epsilon'_{t,d}  &\text{for} ~~ d = 4 ~~ \text{and}~~ t = 251, \dots, 500, \\
 t/T + \epsilon'_{t,d}  & \text{otherwise}, 
\end{cases}  \label{eq:data-setting-1-56}
\end{align}
 where $\epsilon_{t,d} \stackrel{i.i.d.}{\sim} N(0,\sigma^2)$ and $\epsilon'_{t,d} \stackrel{i.i.d.}{\sim} N(0,\sigma^2)$ are independence zero-mean Gaussian noises with variance $\sigma^2 = 0.01$. 
We constructed $X$ and $Y$ so that they differ only in the $4$-th variable, from time index $t =250$ to $500$. 
This information is not known to each method, and the aim is to identify them only from $X$ and $Y$.
Figure~\ref{fig:production-2024-03-12-data-visualization} shows one realisation of $X$ and $Y$. \\

\begin{figure}[t]    
    \centering
    \begin{subfigure}[b]{0.3\textwidth}
        \centering
        \raisebox{0.8cm}{\includegraphics[width=\textwidth]{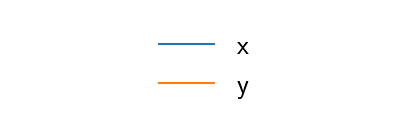}}
    \end{subfigure}
    \hfill
    \begin{subfigure}[b]{0.3\textwidth}
        \centering
        \includegraphics[width=\textwidth]{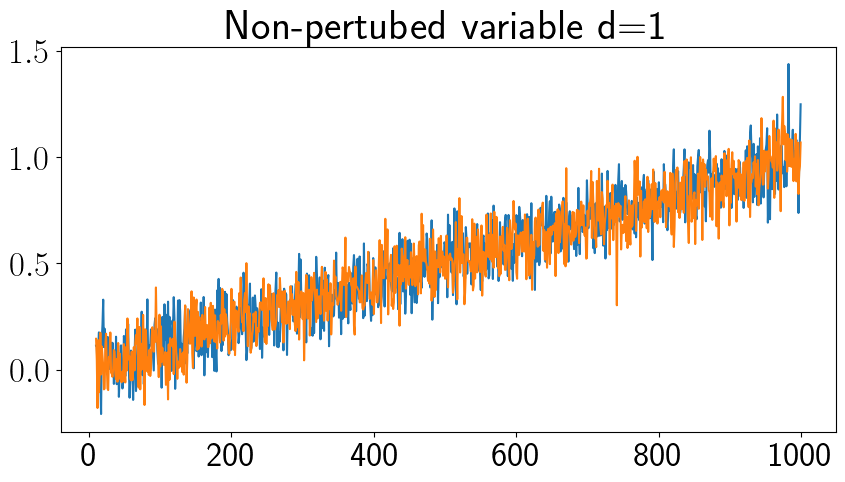}
    \end{subfigure}
    \hfill
    \begin{subfigure}[b]{0.3\textwidth}
        \centering
        \includegraphics[width=\textwidth]{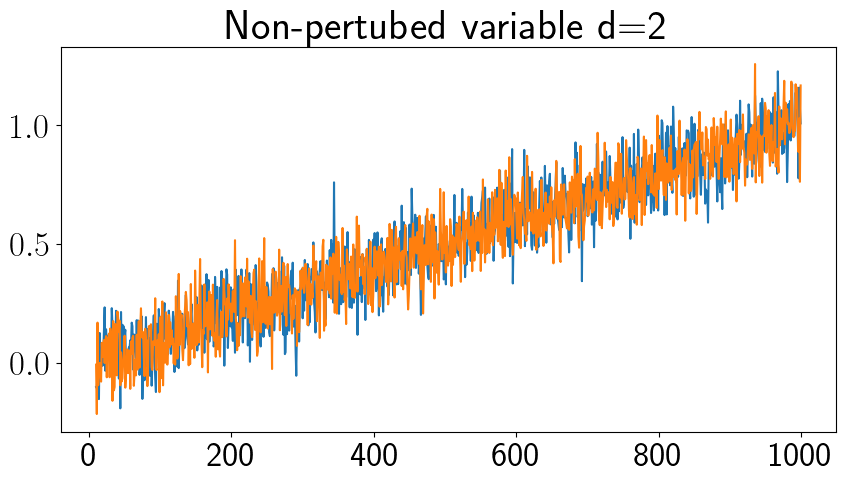}
    \end{subfigure}
    
    \vskip\baselineskip
    
    \begin{subfigure}[b]{0.3\textwidth}
        \centering
        \includegraphics[width=\textwidth]{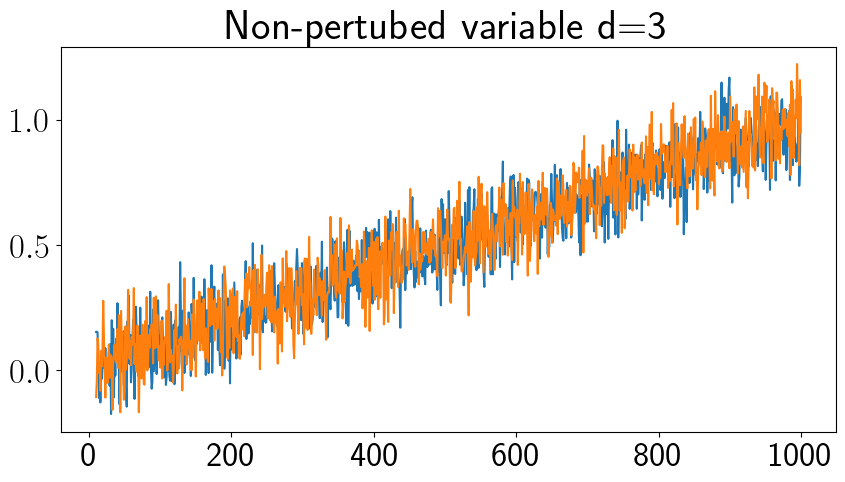}
    \end{subfigure}
    \hfill
    \begin{subfigure}[b]{0.3\textwidth}
        \centering
        \includegraphics[width=\textwidth]{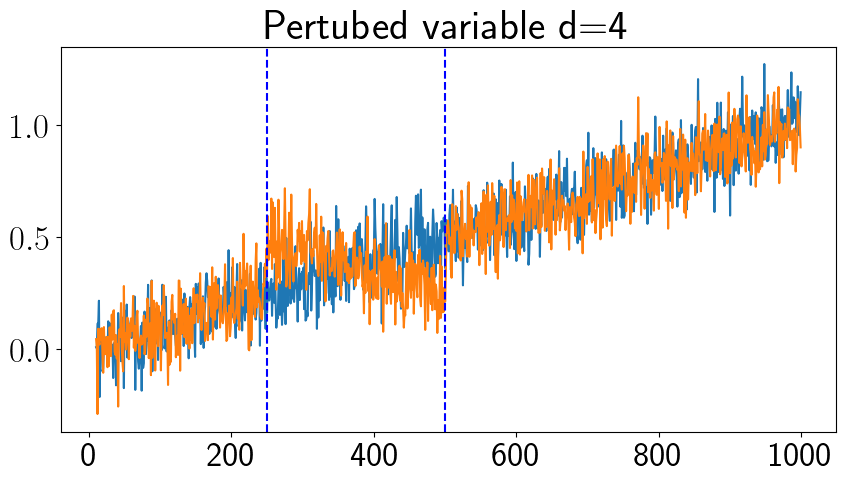}
    \end{subfigure}
    \hfill
    \begin{subfigure}[b]{0.3\textwidth}
        \centering
        \includegraphics[width=\textwidth]{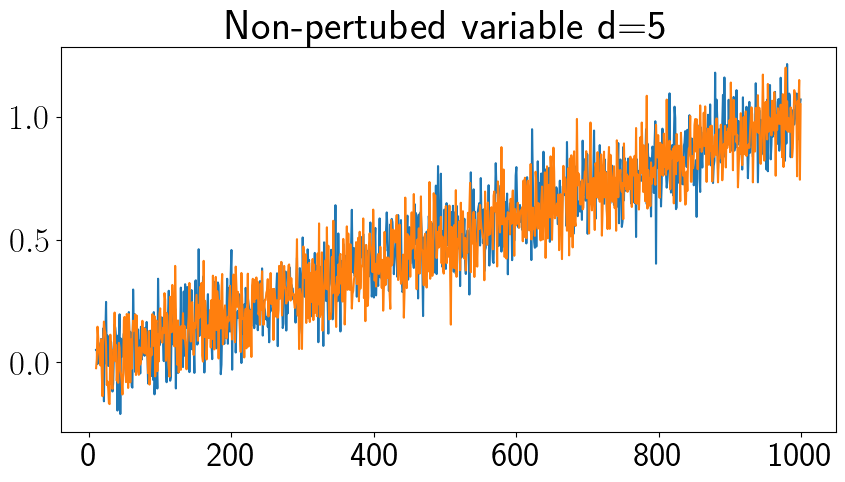}
    \end{subfigure}

    \caption{
    Illustration of two time-series data $X \in \mathbb{R}^{D \times T}$ and $Y  \in \mathbb{R}^{D \times T}$ in  Eq.~\eqref{eq:data-setting-2-316}. 
    Each subfigure shows the trajectories of $X$ and $Y$ in each variable $d= 1,\dots,D$, i.e.,  $x_{1,d}, \dots, x_{T,d}$ and $y_{1,d}, \dots, y_{T,d}$. 
   The variable $d=4$ from $t=251$ to $t = 500$ is where $X$ and $Y$ differ.
    }
    \label{fig:production-2024-03-13-data-visualization}
\end{figure}

\noindent
{\bf Setting 2.}  
We generate $X = (\bx_1, \dots, \bx_T) \in \mathbb{R}^{D \times T}$ and $Y = (\by_1, \dots, \by_T) \in \mathbb{R}^{D \times T}$ as
\begin{align}
x_{t,d} &= t/T + \epsilon_{t,d}   ~~ \text{for} ~~  d = 1,\dots, D ~~ \text{and}~~  t = 1, \dots, T,  \nonumber  \\
y_{t,d} & = 
\begin{cases}
0.5 - (t-250)/T + \epsilon'_{t,d}  &\text{for} ~~ d = 4 ~~ \text{and}~~ t = 251, \dots, 500, \\
 t/T + \epsilon'_{t,d}  & \text{otherwise}, 
\end{cases}  \label{eq:data-setting-2-316}
\end{align}
 where $\epsilon_{t,d} \stackrel{i.i.d.}{\sim} N(0,\sigma^2)$ and $\epsilon'_{t,d} \stackrel{i.i.d.}{\sim} N(0,\sigma^2)$ with $\sigma^2 = 0.01$. 
 See Figure~\ref{fig:production-2024-03-13-data-visualization} for illustration.

Two time-series $X$ and $Y$ differ in the variable $d = 4$ on the period $t = 251, \dots, 500$, i.e., $x_{251, 4}, \dots, x_{500, 4}$ and  $y_{251, 4}, \dots, y_{500, 4}$,  which are the same as the previous setting in Eq.~\eqref{eq:data-setting-1-56}.
The difference from the previous setting is that we generate $y_{251, 4}, \dots, y_{500, 4}$ so that their {\em marginal distribution} becomes the same as $x_{251, 4}, \dots, x_{500, 4}$.
Therefore, if one of subintervals  $[t_{b-1} + 1, t_{b}]$ contains  the period $[251,500]$ entirely, then the proposed framework would fail to detect the change between $x_{251, 4}, \dots, x_{500, 4}$ and  $y_{251, 4}, \dots, y_{500, 4}$.

\subsection{Other Configurations}

{\bf Variable Selection Methods.}
For variable selection in Algorithm~\ref{alg:proposed-framework} (Line 5), we consider the following six methods.
Three methods are based on MMD-based variable selection described in Section~\ref{sec:proposal-mmd-based-variable-selection}: 
(1) \codeName{MMD-Selection},  
(2) \codeName{MMD-CV-AGG}, and
(3) {\tt MMD-Vanilla}, which optimises Eq.~\eqref{eq:mmd-optimisation-problem} without regularisation \citep{sutherland2017}. 
The other three are those based on marginal distribution comparisons in Section~\ref{sec:proposal-wasserstein-based-variable-selection}: (4) {\tt Wasserstein}, which computes the Wasserstein-distances of marginal distributions, (5) \codeName{Mskernel-IMQ}, which computes the MMDs of marginal distributions using the IMQ kernel, and (6) \codeName{Mskernel-Gaussian}, which uses the Gaussian kernel.
As the {\tt Mskernel} methods require the number of variables to be selected as an input, we give the true number of ground-truth variables (which is $1$, as defined above). \\

\noindent
{\bf Time-splitting Points.}
We use equally-spaced time-split points $0 < t_1 < \cdots < t_{B-1} < t_{B} = T$, with two options for the number $B$ of the subintervals: $B = 10$ and $B = 2$. \\

\noindent 
{\bf Train-Test Ratio and Test Statistic.}
In Algorithm~\ref{alg:proposed-framework}, we set $\rho_{\rm train} = 0.8$, and use the sliced Wasserstein distance~\citep{bonneel-2015} as a test statistic in the permutation test statistics for each subinterval.\\

\noindent
{\bf Evaluation Metrics.}
To evaluate each variable selection method applied to each subinterval, we compute 
$
{\rm Precision} = |\VariableSelectionHat \cap S| / |\VariableSelectionHat|$ and 
${\rm Recall} = |\VariableSelectionHat \cap S| / |S| 
$,
where $S \subset \{1,\dots, D\}$ is the set of ground-truth variables and $\VariableSelectionHat  \subset \{1,\dots, D\}$ is the set obtained by the variable selection method.

\begin{figure*}[t!]
    \captionsetup[subfigure]{labelformat=empty}
    \centering
    \includegraphics[scale=0.4]{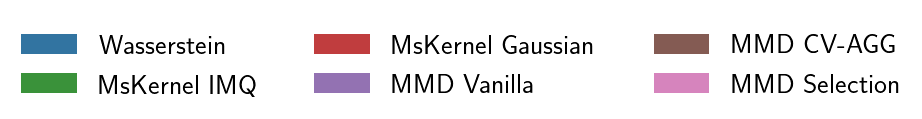}
    \hfill
    \begin{minipage}{.4\textwidth}
        \includegraphics[width=\textwidth]{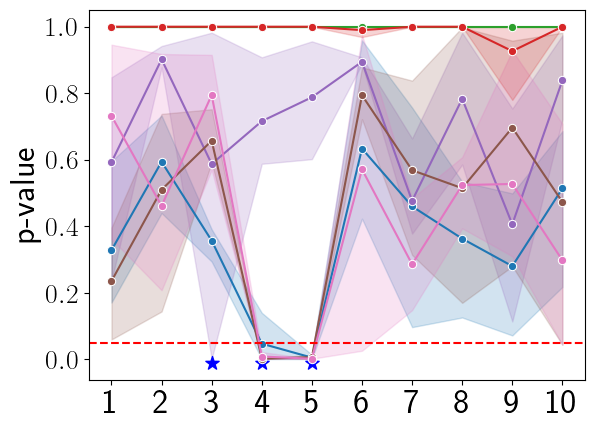}
    \end{minipage} ~~ 
    \begin{minipage}{.40\textwidth}
        \includegraphics[width=\textwidth]{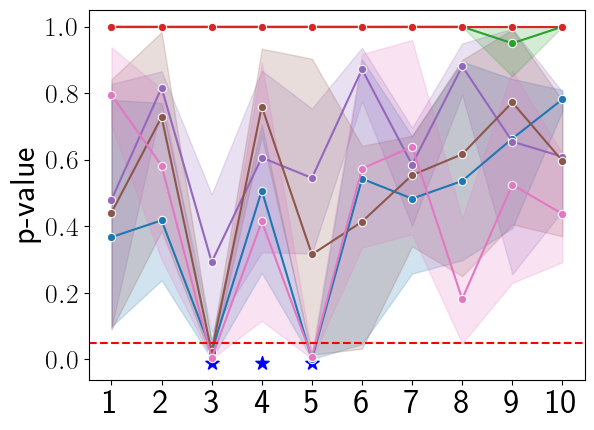}
    \end{minipage} ~~ 
    \begin{minipage}{.40\textwidth}
        \includegraphics[width=\textwidth]{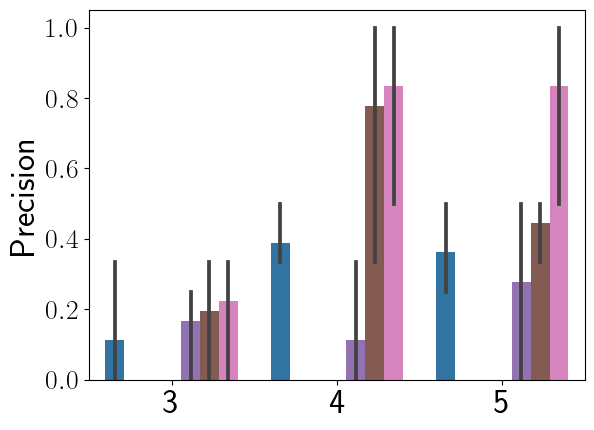}
    \end{minipage} ~~ 
    \begin{minipage}{.40\textwidth}
        \includegraphics[width=\textwidth]{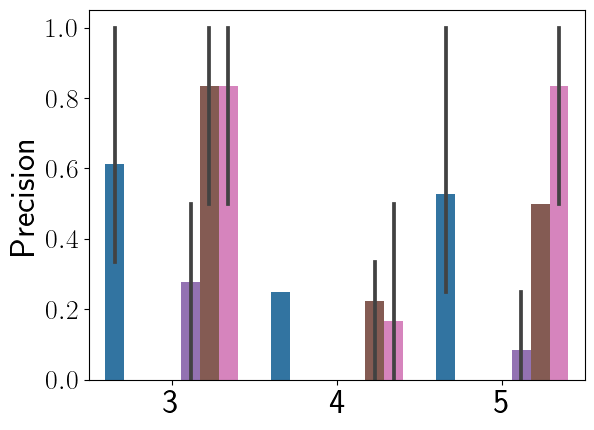}
    \end{minipage} ~~ 
    \begin{minipage}{.40\textwidth}
        \includegraphics[width=\textwidth]{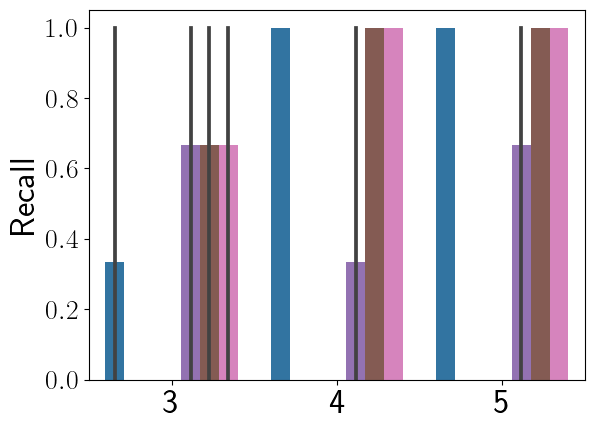}
               \subcaption{\quad \quad   Setting 1 with $B =10$.}
    \end{minipage} ~~ 
    \begin{minipage}{.40\textwidth}
        \includegraphics[width=\textwidth]{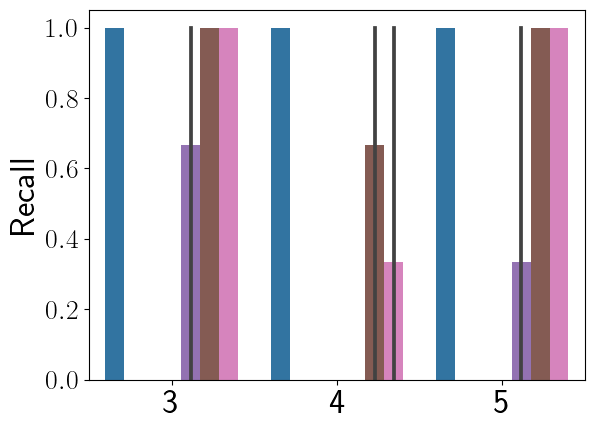}
    
        \subcaption{\quad \quad  Setting 2 with $B = 10$.}
    \end{minipage} 

    \caption{
        Results of the experiments in Section~\ref{sec:synthetic-exp} with $B=10$ in Setting 1 (left column) and  Setting 2 (right column). 
        {\bf Top}: The p-values over the $B=10$ subintervals obtained by each method, where each line and the shaded area represent the means and standard deviations computed over three independently realisations of $X$ and $Y$. 
        The red dotted horizontal line indicates the value $0.05$.
        The three blue stars indicate the three subintervals, 3, 4 and 5, that intersect with the changing period $[251,500]$. 
        {\bf Middle}:
        The precision scores for each method, where the bars and error bars represent the means and standard deviations over three independent realisations of $X$ and $Y$. 
        The numbers on the horizontal axis (3, 4 and 5) indicate the  subintervals that intersect the changing period $[251,500]$.    
        {\bf Bottom}:
        The corresponding recall sores. 
                Note that the precision and recall for {\tt MsKernel-Gaussian}, {\tt MsKernel-IMQ} and {\tt MMD-Vanilla} were zero, so they are not shown. 
        }
    \label{fig:exp-one-results}
\end{figure*}

\begin{figure}[t]
    \centering
    \captionsetup[subfigure]{labelformat=empty}
    \includegraphics[scale=0.4]{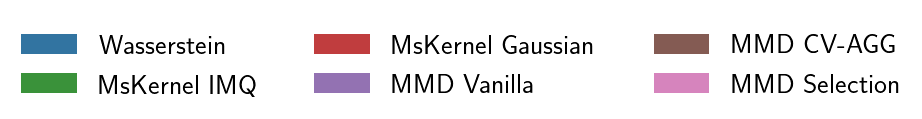}
    \hfill
    \begin{minipage}{\textwidth}
        \centering
        \subcaption{\quad \quad P-values for $B = 2$}
        \includegraphics[scale=0.6]{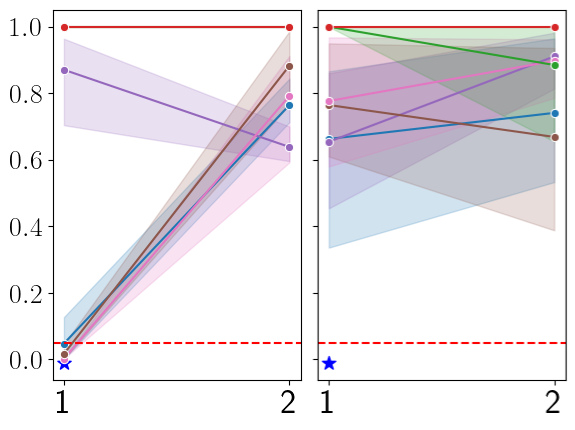}
        \subcaption{\quad \quad  Left: Setting 1.\quad Right: Setting 2.}
    \end{minipage}
    \begin{minipage}{.40\textwidth}
        \centering
       \subcaption{\quad \quad  Precision for $B = 2$} 
        \includegraphics[width=\textwidth]{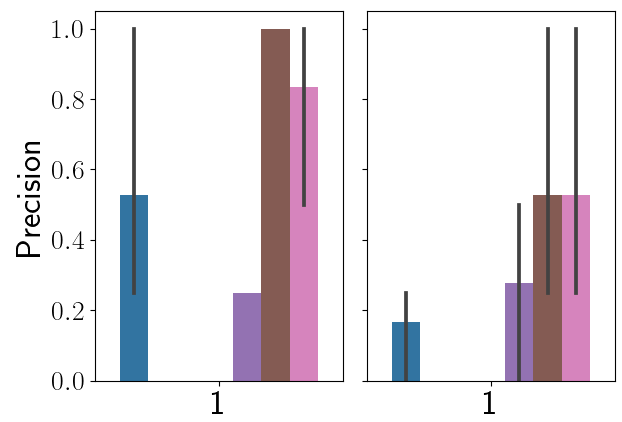}
        \subcaption{\quad \quad  Left: Setting 1.\quad Right: Setting 2.}
    \end{minipage}~~ 
    \begin{minipage}{.40\textwidth}
        \centering
        \subcaption{\quad \quad  Recall for $B = 2$}
        \includegraphics[width=\textwidth]{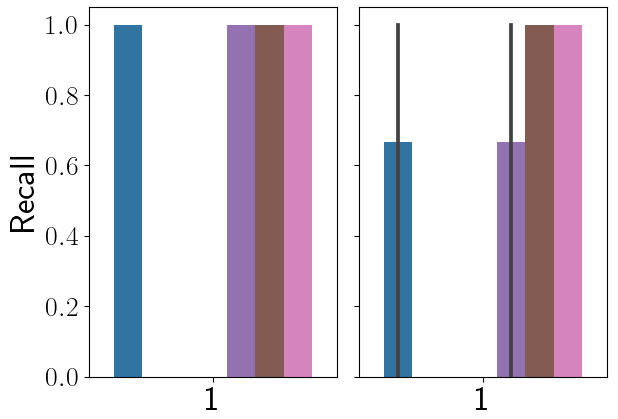}
     \subcaption{\quad \quad  Left: Setting 1.\quad Right: Setting 2.}
    \end{minipage}\hfill

    \caption{
     Results of the experiments in Section~\ref{sec:synthetic-exp} with $B=2$. 
     In each figure, the left and right subfigures are the results for Settings 1 and 2, respectively. 
     The top figure shows the p-values, the bottom left the precision scores, and the bottom right the recall scores.  
     For details, see the caption of Figure~\ref{fig:exp-one-results}.
        }
        \label{fig:exp-two-results}
\end{figure}

\subsection{Results}
\label{sec:experiment-result}
For ease of comparison, we summarise the results for $B = 10$ in Figure~\ref{fig:exp-one-results} and $B =2$ in Figure~\ref{fig:exp-two-results}. 
Each of Figures~\ref{fig:exp-one-results} and \ref{fig:exp-two-results} shows the means and standard deviations of p-values, precision, and recall over three independent experiments for each of Settings 1 and 2.
Precision and recall scores are shown only for subintervals overlapping the period $[251, 500]$ where changes in $X$ and $Y$ occur (as the recall is not well-defined for the other subintervals where there exists no ground-truth variables $S$).
{\tt MMD-Vanilla}, {\tt MsKernel-Gaussian} and {\tt MsKernel-IMQ} failed to identiy the differing variable $d = 4$ on the 3rd, 4th and 5th intervals in most settings, leading to zero precision and zero recall. 
We thus focus on discussing the other three methods.

Let us first study the case $B = 10$ in Figure~\ref{fig:exp-one-results}, where each subinterval consists of $100$ points.
The 3rd ($[201,300]$), 4th ($[301,400]$) and 5th ($[401,500]$) intervals overlap the period $[251,500]$ where $X$ and $Y$ differ.  
We can make the following observations.
\begin{itemize}
    \item In Setting 1, {\tt MMD-CV-AGG}, {\tt MMD-Selection} and {\tt Wasserstein} gave low p-values on intervals 4 and 5, suggesting that these methods selected the correct variable ($d=4$). (Recall that all methods use the same test statistic for the permutation test, so they only differ in the select variables.)  
    Indeed, the recall is $1$ for these methods on intervals 4 and 5:  the correct variable is included in their selected variables. 
    On the other hand, the precision is about $0.8$ for {\tt MMD-CV-AGG} and {\tt MMD-Selection} and about $0.4$ for {\tt Wasserstein}, suggesting that {\tt Wasserstein} selected more redundant variables. 

    \item On interval 3 in Setting 1, the p-values are not small for these methods. The precision and recall are also lower than intervals 4 and 5. 
    This is because variable selection is more difficult for interval 3.
    Indeed, the intersection of  interval 3 and the differing period $[251,500]$ is $[251, 300]$, which contains only 50 points and half of those for intervals 4 and 5.
    Moreover, as can be seen from Figure~\ref{fig:production-2024-03-12-data-visualization} ($d=4$), the difference between $X$ and $Y$ appears only slightly on interval 3.

    \item In Setting 2, the p-values are small on intervals on 3 and 5 for {\tt MMD-CV-AGG}, {\tt MMD-Select} and {\tt Wasserstein}. 
    As the recall is $1$ and the precision is from $0.5$ to $0.8$ for these methods, they were able to select the correct variable $d = 4$ on these intervals. As can be seen in Figure~\ref{fig:production-2024-03-13-data-visualization} ($d=4$), the difference between $X$ and $Y$ is clear on these intervals, so correct variable selection was also possible for interval 3. 

    \item On the other hand, interval 4 is more difficult for Setting 2, as the difference between $X$ and $Y$ is more subtle, thus producing large p-values. (The sample size of $100$ may not be sufficient to identify the difference).

    \item Producing large p-values outside intervals 3, 4 and 5 is correct, as there is no difference between the generating processes of $X$ and $Y$. 
\end{itemize}
These observations suggest that Algorithm~\ref{alg:proposed-framework}, using {\tt MMD-CV-AGG}, {\tt MMD-Select} or {\tt Wasserstein}, can perform correct variable selection (at least in the considered settings) on the subintervals intersecting with the changing period and thus can detect such a period, if the differences between $X$ and $Y $ are clear enough for given sample sizes.

Now let us see the case $B = 2$ in Figure~\ref{fig:exp-two-results}, where the entire interval is divided into two subintervals $[1, 500]$ and $[501, 1000]$, each consisting of $500$ points.
The changing period $[251,500]$ is contained in the first interval $[1, 500]$. 
The following can be observed:
\begin{itemize}
    \item In Setting 1, the p-values are small for {\tt MMD-CV-AGG}, {\tt MMD-Selection} and {\tt Wasserstein} on interval 1, where the recall is $1$ and the precision is higher than $0.5$ for these methods. 
    Since interval 1 contains the changing period $[251,500]$, this suggests that these methods were able to select the correct variable $d = 4$.

    \item In Setting 2, the p-values for interval 1 are large for all the methods, suggesting the failure of variable selection. Indeed, the precision of each method is lower than Setting 1.
    As can be seen from Figure~\ref{fig:production-2024-03-13-data-visualization} ($d=4$), if we ignore the time order of points on interval 1 (which is essentially done by Algorithm~\ref{alg:proposed-framework}), the marginal distributions for $d=4$ are (nearly) identical for $X$ and $Y$ by construction. Therefore, one cannot detect the difference between $X$ and $Y$ on interval 1 by just looking at the marginal distributions for $d = 4$, so variable selection is harder than Setting 1.  
    (Compare this situation with the case of $B = 10$, where variable selection was possible even in Setting 2.) 
    \item However, variable $d = 4$ is correlated with other variables, and this correlation structure differs for $X$ and $Y$. Therefore, it would still be possible to select variable $d=4$ by identifying the change in the correlation structure. Indeed, the precision of {\tt MMD-CV-AGG} and {\tt MMD-Selection} is about $0.5$, so they could identify the change for $d = 4$ to some extent. On the other hand, the precision of {\tt Wasserstein} is about $0.2$, which is the level of precision achievable by selecting all the five variables. This is unsurprising as the algorithm of {\tt Wasserstein}, as designed here, does not consider the correlation structure among variables. 
\end{itemize}
We can summarise the above observations as follows: If we make the number of subintervals $B$ too small, each subinterval becomes longer, making it harder to detect changing variables.  
Therefore, it is important to make  $B$ not too small so that each subinterval is short enough to detect the changes.
It is also shown that {\tt MMD-CV-AGG} and {\tt MMD-Selection} perform better than the other methods used in the experiments.

\section{Demonstration: Validation of a DNN Model for Particle-based Fluid Simulation}

\label{sec:demonstration}

\label{sec:demonstration-particle}

We demonstrate how our approach can be used to validate a Deep Neural Network (DNN) model that emulates a particle-based fluid simulator. Motivated by the need for accelerating computationally expensive fluid simulations, there has been a recent surge of interest in developing DNN models that learn to approximate fluid simulations~\citep[e.g.,][]{ummehofer-2020, sanchez-2020, prantl-2022}. However,  model validation of a DNN emulator against the original simulator remains challenging, as both produce high-dimensional spatio-temporal outputs. Here, we demonstrate that our approach may be used to produce interpretable features that could be used by the human modeller for validation purposes.

\subsection{Setup}
\label{sec:demonstration-particle-setup}

\begin{figure}[t]
    \centering
    \includegraphics[width=0.45\linewidth]{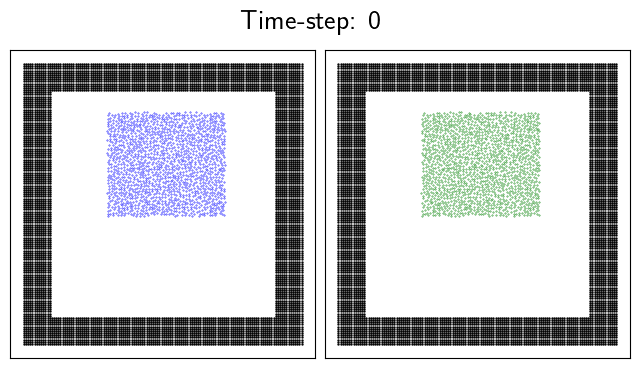}
    \includegraphics[width=0.45\linewidth]{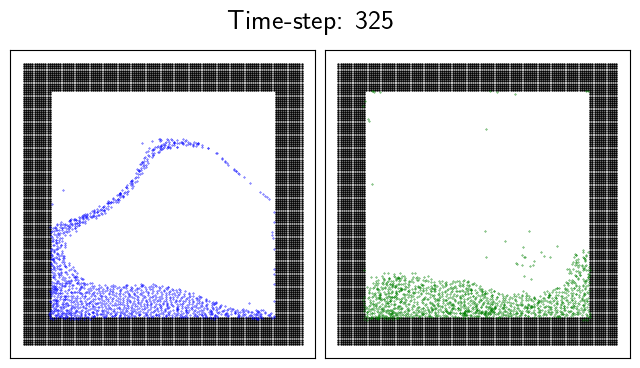}
    \caption{
    Snapshots of water particle distributions of the {\tt WaterRamp} scenario in Section~\ref{sec:demonstration-particle-setup}, at time $t = 0$ (left two figs) and time $t = 325$ (right two figs). 
    The blue particles are from {\tt Splish-Splash}, the ground-truth simulator, while the green ones are from the learned {\tt DMCF} model. 
    }
      \label{fig:demonstration-particle-snapshot-pickup}
\end{figure}

\noindent
{\bf Simulator.}
The ground-truth simulator is {\tt Splish-Splash}~\citep{SPlisHSPlasH_Library}, an open-source particle-based fluid simulator.
Specifically, we consider the {\tt WaterRamp} scenario from \citet{sanchez-2020}, which simulates water particles in a box environment, as illustrated in Figure~\ref{fig:demonstration-particle-snapshot-pickup}. \\

\noindent
{\bf DNN model.}
As the DNN model for learning the simulator, we use the {\tt Deep Momentum-Conserving Fluids (DMCF)} model  of \citet{prantl-2022}, using the authors' code\footnote{\url{https://github.com/tum-pbs/DMCF/blob/main/models/cconv.py}}, trained with the default configuration on the dataset provided by the authors.\footnote{\url{https://github.com/tum-pbs/DMCF/blob/96eb7fcdd5f5e3bdda5d02a7f97dfff86a036cfd/download_waterramps.sh}}  \\

\noindent
{\bf Data Representation.}
Each of the simulator and the DNN model produces an output consisting of an array in $\mathbb{R}^{P \times \TimeN \times C}$, 
where $P=1,444$ is the number of particles\footnote{The number of particles varies depending on the simulation random seed, which we set to $0$.},  $T = 600$ is the number of time steps, and $C=2$ is the number of coordinates for each particle's position (i.e., two dimensions). 
We convert this array to a matrix in $\mathbb{R}^{D \times T}$, by dividing the two-dimensional coordinate space into $D = 256 = 16 \times 16$ grids of equal-size squares.
Thus, for a given output $X = (x_{t,d}) \in \mathbb{R}^{D \times T}$, each $x_{t,d}$ represents the number of particles on the $d$-th grid at the $t$-th time step, where $d = 1,\dots, D$ and $t = 1,\dots, T$. 
Intuitively, each grid represents a ``sensor'' that counts the number of particles that passes the grid. \\

\noindent
{\bf Setting of Algorithm~\ref{alg:proposed-framework}.}
We split the $T = 600$ time steps into $B = 6$ subintervals, each consisting of $100$ steps: $t_1 = 100, t_2 = 200, \dots, t_6 = 600$.
For variable selection in Algorithm~\ref{alg:proposed-framework}, we consider three methods:  \codeName{Wassertstein},  \codeName{MMD-Selection}, and \codeName{MMD-CV-AGG}, as they performed better for the experiments in Section~\ref{sec:experiment}.
We use the same configurations as Section~\ref{sec:experiment-mmd-configuration} for these methods, with one modification. For the MMD-based methods, we use the dimension-wise {\em mean} (instead of median) heuristic \citep[Appendix C]{mitsuzawa-2023} to set the length scales $\gamma_1, \dots, \gamma_D$ of the kernel \eqref{eq:ARD-kernel}, as this leads to more stable results when the data is sparse, i.e., having many zeros, like the current setting.

\subsection{Results}
\label{sec:demonstration-particle-result}

\begin{figure}[htbp]
    \centering
    \begin{subfigure}[b]{0.45\textwidth}
        \centering
        \includegraphics[width=\textwidth]{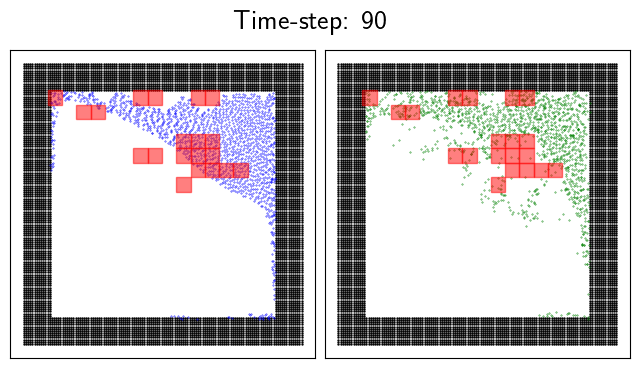}
        \caption{{\tt Wasserstein} (1st subinterval $[1,100]$)}
        \label{fig:subfig1}
    \end{subfigure}
    \hfill
    \begin{subfigure}[b]{0.45\textwidth}
        \centering
        \includegraphics[width=\textwidth]{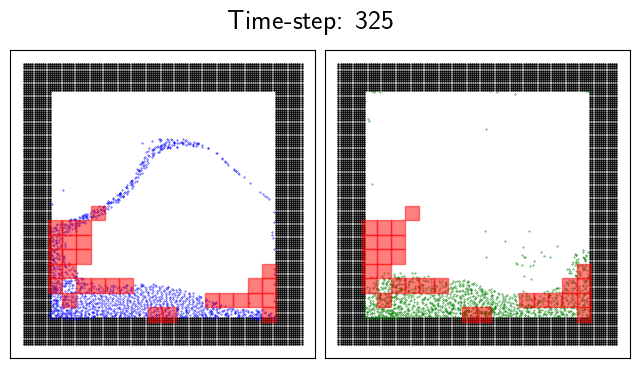} 
        \caption{{\tt Wasserstein} (4th subinterval $[301,400]$)}
        \label{fig:subfig2}
    \end{subfigure}
    
    \noindent\rule{\textwidth}{0.4pt} 
    
    \begin{subfigure}[b]{0.45\textwidth}
        \centering
        \includegraphics[width=\textwidth] {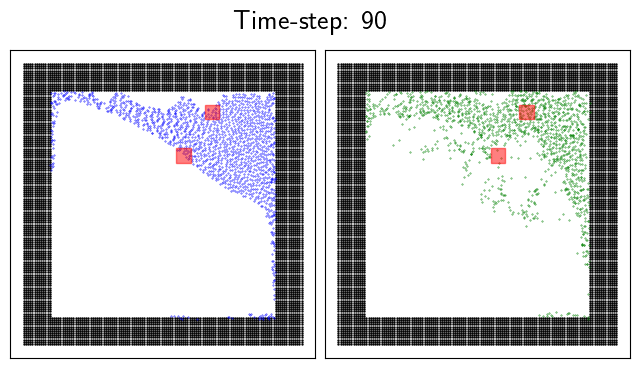} 
        \caption{ {\tt MMD-Selection} (1st subinterval $[1,100]$)}
        \label{fig:subfig3}
    \end{subfigure}
    \hfill
    \begin{subfigure}[b]{0.45\textwidth}
        \centering
        \includegraphics[width=\textwidth]{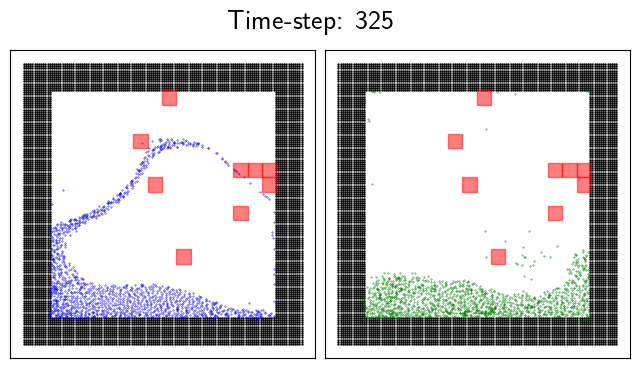} 
        \caption{ {\tt MMD-Selection} (4th subinterval $[301,400]$)}
        \label{fig:subfig4}
    \end{subfigure}
    
    \noindent\rule{\textwidth}{0.4pt} 
    
    \begin{subfigure}[b]{0.45\textwidth}
        \centering
        \includegraphics[width=\textwidth]{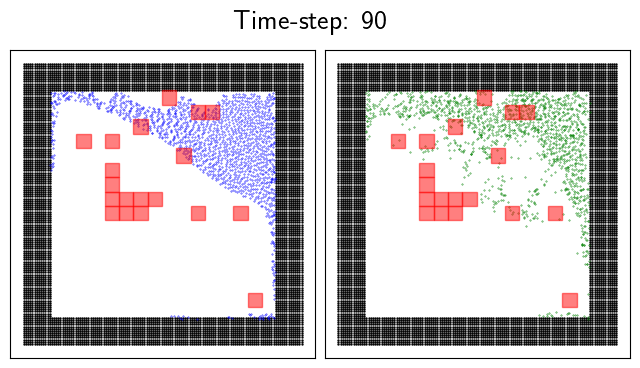} 
        \caption{{\tt MMD-CV-AGG} (1st subinterval $[1,100]$)}
        \label{fig:subfig5}
    \end{subfigure}
    \hfill
    \begin{subfigure}[b]{0.45\textwidth}
        \centering
        \includegraphics[width=\textwidth]{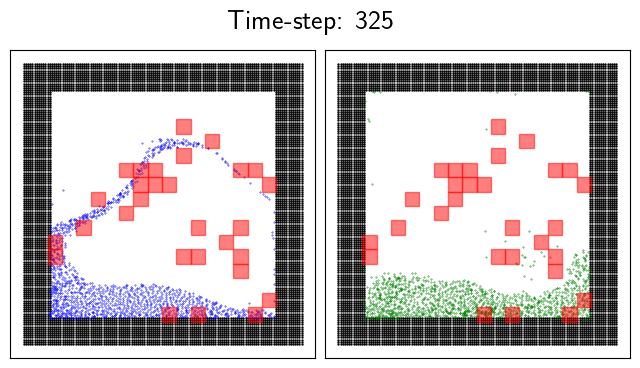}
        \caption{{\tt MMD-CV-AGG} (4th subinterval $[301,400]$)}
        \label{fig:subfig6-MMD-CV-AGG}
    \end{subfigure}
    
    \caption{   
    Selected results from  the experiments in Section~\ref{sec:demonstration-particle-result}.
    The top row shows the selected variables using
     Algorithm~\ref{alg:proposed-framework} with {\tt Wasserstein}, the middle row with {\tt MMD-Selection}, and the bottom row with {\tt MMD-CV-AGG}. 
     In each row, the left two figures show the selected variables for the 1st subinterval $[1, 100]$, and the right two figures for the 4th subinterval $[301,400]$.
    The red grids represent the selected variables, the blue particles are those from the ground-truth simulator and the green ones are from the DNN model (at time $t = 90$ for the left and $t = 325$ for the right). }
    \label{fig:particle-sim-selected-vars}
\end{figure}

    \begin{figure}[t]
       \centering
        \includegraphics[width=0.50\linewidth]{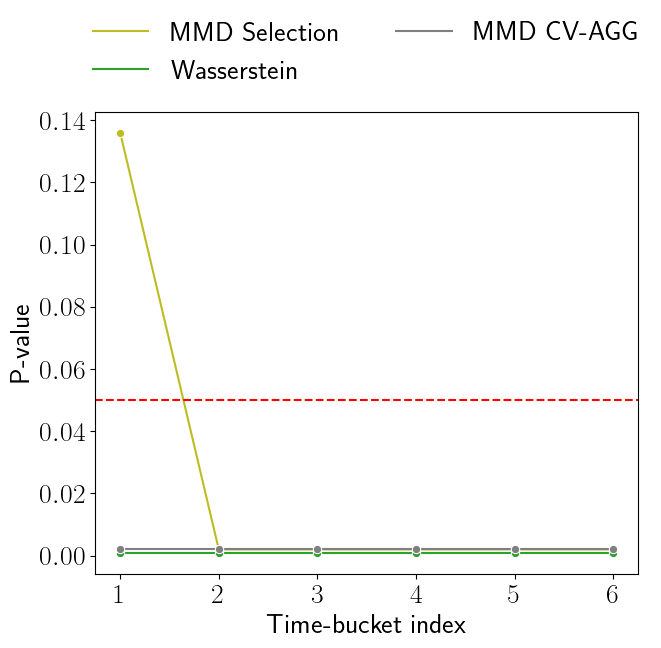}
        \caption{
            P-values obtained with Algorithm~\ref{alg:proposed-framework} on each subinterval with the  {\tt MMD-Selection}, {\tt MMD-CV-AGG} and {\tt Wasserstein} for the experiments in Section~\ref{sec:demonstration-particle}.
            ``Time-bucket index'' in the horizontal axis indicates the six subintervals. 
            The red dashed line indicates the value $0.05$.
        }
        \label{fig:demonstration-particle-sensor-256-p-values}
    \end{figure}

Figure~\ref{fig:particle-sim-selected-vars} shows selected results of variable selection.
More informative GIF animations illustrating selected variables are available on the authors' website\footnote{\url{https://kensuke-mitsuzawa.github.io/\#/publications/mmd-paper-demo-particle-simulation}}.
Figure~\ref{fig:demonstration-particle-sensor-256-p-values} shows p-values obtained with each variable selection method on each subinterval. 
We can make the following observations:
\begin{itemize}
    \item Generally, selected variables (= red-highlighted grids) can indicate the discrepancies between the ground-truth simulator and the DNN model.  
    For example, let us look at the grids selected by {\tt MMD-CV-AGG} on the 4th interval $[301, 400]$  (Figure~\ref{fig:subfig6-MMD-CV-AGG}).  
    Many grids are selected along the blue particles' wave-like curve formed in the middle of the box (the ground-truth).  
    However, such a wave-like curve does not exist for the green particles from the DNN model.  
    Thus these selected grids successfully capture this  discrepancy between the simulator and the DNN model.

    \item There are differences between the variables selected by the three methods. 
    For example, {\tt Wasserstein} and {\tt MMD-Selection} only partially capture the above discrepancy regarding the wave-like curve. 
    On the other hand, {\tt MMD-Selection} selects one grid along the ceiling of the box, which captures another discrepancy between the simulator and the DNN model: While there is no particle from the simulator, there are a few particles from the DNN model sticking on the ceiling, which are not physically plausible.

    \item Let us look at Figure~\ref{fig:demonstration-particle-sensor-256-p-values} on p-values.
On the 2nd to 6th subintervals, the p-values are all less than 0.05 for each method, indicating that the simulator and the DNN model are significantly different on these subintervals. 
On the 1st subinterval $[1,100]$, {\tt MMD-CV-AGG} and {\tt Wasserstein} lead to p-values less than 0.05, while {\tt MMD-Selection} yields a p-value higher than $0.05$. 
The latter would be because {\tt MMD-Selection} selects only two variables (the two red grids in Figure~\ref{fig:particle-sim-selected-vars}), which may not be enough for capturing the discrepancies between the two datasets on the 1st interval. ({\tt MMD-Selection} tends to select few variables than {\tt MMD-CV-AGG} in general.)

\item On the other hand, the fact that {\tt MMD-Selection} gives a larger p-value on the first subinterval suggests that the discrepancies between the simulator and the DNN model are less significant compared to other subintervals. This is indeed the case, as the initial states are the same for the simulator and the DNN model.
\end{itemize}
From these observations, when using Algorithm~\ref{alg:proposed-framework} in practice, we recommend the user to try different variable selection methods (if possible) and compare the results. In this way, a more thorough analysis becomes possible.

\section{Demonstration: Comparison of Microscopic Traffic Simulations}
\label{sec:demonstration-most}

We demonstrate how the proposed approach can be used to analyse the effects of changing a simulator's configuration on its output, using a microscopic traffic simulator as an illustrative example.  
Such an analysis is hard to be done manually, as a traffic simulator produces high-dimensional time-series data. 
The proposed approach is helpful to aid a human analyst by providing variables and time intervals where the effects of changing the configuration are prominent.

Supplementary results, including animations of simulations and variable selection results, are available on the authors' website.\footnote{\url{https://kensuke-mitsuzawa.github.io/\#/publications/mmd-paper-demo-sumo-most}}

\subsection{Setup}
\label{sec:demonstration-most-setup}

\begin{figure}[t]
\centering 
    \includegraphics[width=0.7 \linewidth]{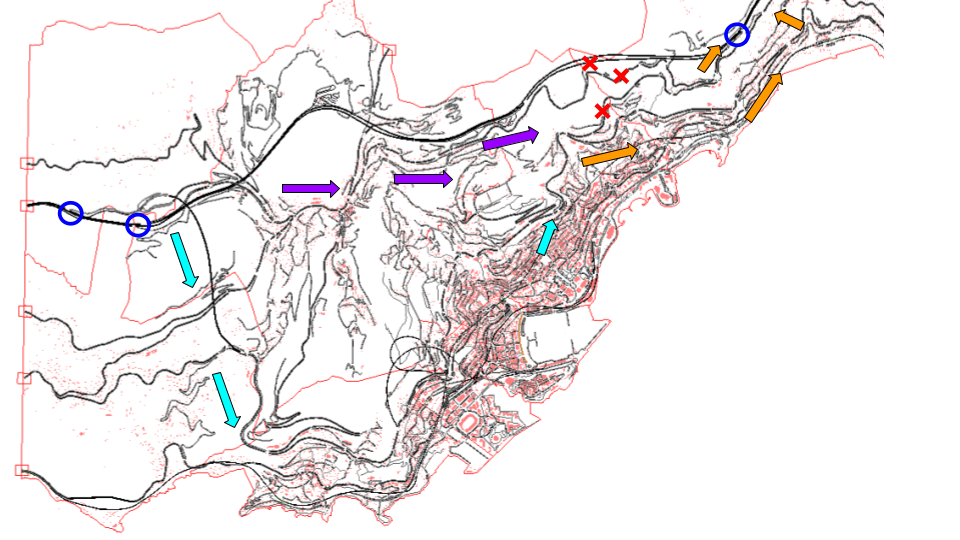}
    \caption{
        The simulated area of the {\tt MoST} scenario inside and around Monaco in Section~\ref{sec:demonstration-most}.
        The three red ``X" marks indicate the blocked locations in the {\tt A8} highway (top), {\tt D2564} (middle) and {\tt D51} (bottom), in the modified scenario.
        The blue circles indicate ramps to the {\tt A8} highway.  
        The purple and light-value arrows indicate the rerouting paths of vehicles that originally planned to travel via the {\tt A8} highway. 
    }
    \label{fig:demonstration-most-study-map}
\end{figure}

We use {\tt SUMO}~\citep{SUMO2018}, a popular microscopic traffic simulator. 
We consider the simulation scenario named {\tt MoST} (\textbf{Mo}naco \textbf{S}umo \textbf{T}raffic) developed by \citet{codeca-2018}.
It models a realistic, large-scale traffic scenario inside the Principality of Monaco and the surrounding area on the French Riviera, as described in Figure~\ref{fig:demonstration-most-study-map}. 
The scenario simulates various modes of transportation, such as passenger cars, public buses, trains, commercial vehicles, and pedestrians.
Here, we focus on the simulation from 4 a.m.~until 2 p.m., the duration being 6 hours (= 36,000 seconds).\footnote{
In the {\tt SUMO} implementation of the {\tt MoST} scenario, this duration is from the 14,400th step to the 50,400th step, where one step corresponds to one second in the real world.}
\\

\noindent
{\bf Simulation Scenarios.}
We consider two versions of the {\tt MoST} scenario, one original and the other a modified version.  
We define the modified version\footnote{This modified scenario is available on \url{https://github.com/Kensuke-Mitsuzawa/sumo-sim-monaco-scenario}} by blocking three roads, {\tt A8} (highway), {\tt D2564} and {\tt D51}, in the original {\tt MoST} scenario, at the locations marked ``X'' in red in Figure~\ref{fig:demonstration-most-study-map}.
This road blocking affects the vehicles that originally planned to use the {\tt A8} highway (where the top ``X'' blocks) to change their routes. 
One group of vehicles travel on the roads next to the {\tt A8} highway, as indicated by the purple arrows in Figure~\ref{fig:demonstration-most-study-map}.  
As light-blue arrows indicate, another group changes their route to the south and passes inside Monaco, making the traffic there heavier than the original scenario.
These two groups of vehicles meet again at a junction\footnote{The geographical coordinate of this junction is $(43.7522398630394, 7.425510223260723)$.} near the bottom ``X'' and travel back to the {\tt A8} highway, as indicated by orange arrows.\footnote{The vehicles in the opposite direction (towards the west) also change their routes similarly.}

Note that the above observations about the influences of the blocked roads are not known beforehand; they are obtained through manual data analysis based on the insights given by our variable selection approach. 
Therefore, they are post-hoc explanations, but we present them here for ease of understanding. 
\\

\noindent
{\bf Data Representation.}
We simulate both scenarios using the same random seed.\footnote{We set the random seed value to 42.}
Let $D = 4,404$ be the number of road segments\footnote{While ``edge" is the proper {\tt SUMO} terminology, we call it ``road segment'' for ease of understanding.} in the simulation area, and $T = 3,600$ be the number of time steps from 4 a.m.~to 2 p.m.~(= 36,000 seconds) where each time step is 10 seconds.\footnote{Because each step in the {\tt SUMO} simulator corresponds to one second, this means that we aggregate each 10 {\tt SUMO} steps to form one time step in our data representation.} 
For the original scenario, let  $x_{d,t}$ be the number of vehicles on the road segment $d = 1,\dots,D$ during the time step $t = 1, \dots, T$, and define the data matrix as $X = (x_{t,d}) \in \mathbb{R}^{D \times T}$. 
For the modified scenario, we construct $Y = (y_{t,d}) \in \mathbb{R}^{D \times T}$ similarly.\\

\noindent
{\bf Variable Selection Methods.}
For variable selection in Algorithm~\ref{alg:proposed-framework}, we try {\tt Wasserstein},  \codeName{MMD-Selection} and \codeName{MMD-CV-AGG}, with the same configurations as Section~\ref{sec:demonstration-particle}.
We set the time-splitting points as $t_1 = 500$, $t_2 = 1,000$, ... $t_{7} = 3,500$ and $t_8 = 3,600$, resulting in $B = 8$ subintervals.

\subsection{Results}
\label{sec:demonstration-most-result}

\begin{figure}[t]
    \begin{minipage}{0.45\textwidth}
        \includegraphics[width=\textwidth]{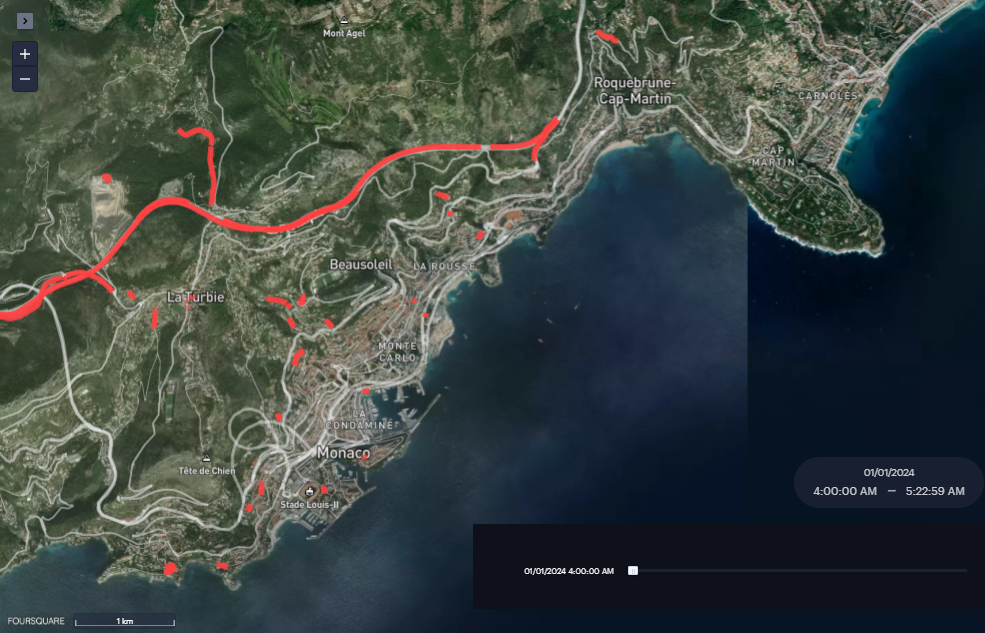}
        \subcaption{Selected road segments on interval 1 (4:00 am - 5:23 am).}
    \end{minipage} \hfill
    \begin{minipage}{0.45\textwidth}
        \includegraphics[width=\textwidth]{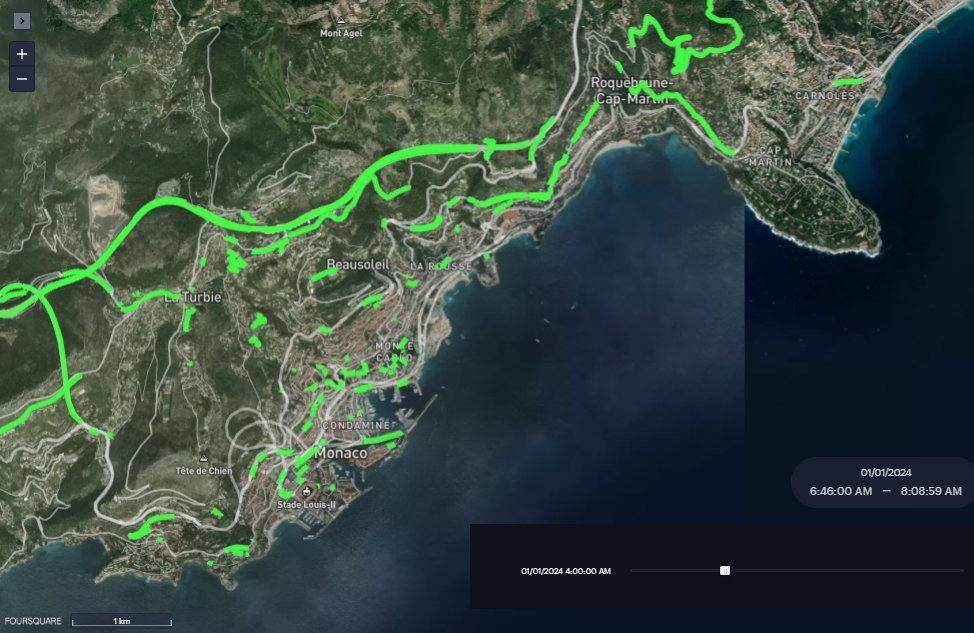}
        \subcaption{Selected road segments on interval 3 (6:46 am - 8:09 am.}
    \end{minipage}
    \\
    \begin{center}
        \begin{minipage}{0.45\textwidth}
            \includegraphics[width=\textwidth]{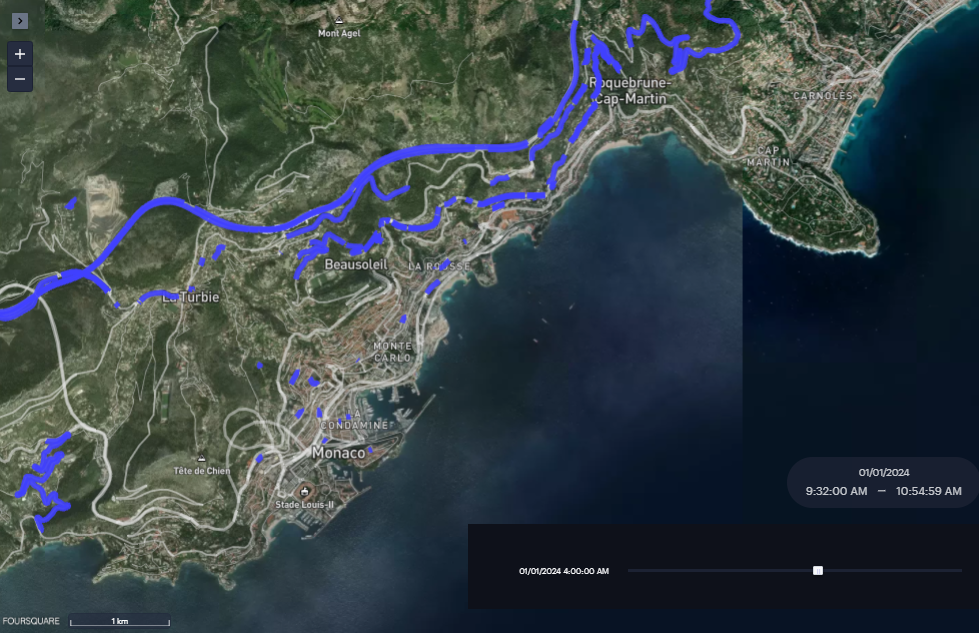}
            \subcaption{Selected road segments on interval 5 (9:32 am - 10:55 am).}
        \end{minipage}
    \end{center}
    \caption{
        Road segments selected by \codeName{MMD-CV-AGG} on the 1st (a), 3rd (b) and 5th (c) subintervals in the experiments of Section~\ref{sec:demonstration-most}.
    }
    \label{fig:demonstration-most-pickup-by-mmd-cv-overview}
\end{figure}

\begin{figure}[t]
    \includegraphics[width=.45\textwidth]{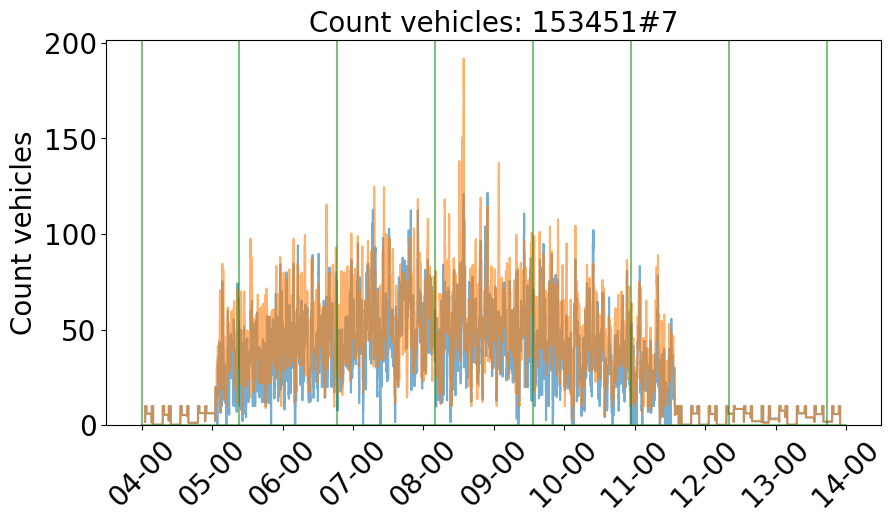} \hfill
    \includegraphics[width=.45\textwidth]{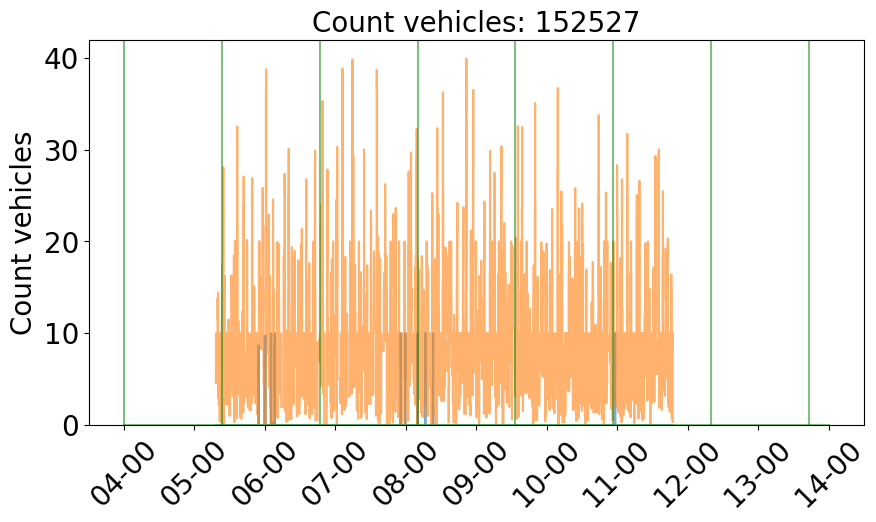} \hfill
    \caption{
        Time-series $x_{1,d}, \dots, x_{T,D}$ from the original scenario (blue) and $y_{1,d}, \dots, y_{T,D}$ from the modified scenario (orange) for two specific road segments $d$ in Section~\ref{sec:demonstration-most}. 
        The left figure is where the road segment $d$ is at the centre of commune \textit{La Turbie} (road id \codeName{153451\#7}; near the left most purple arrow in Figure~\ref{fig:demonstration-most-study-map}).
        The right figure is where  $d$ is at an entrance ramp to the {\tt A8} highway and is located at the north part of commune \textit{Dondéa} (road id \codeName{152527}; near the blue circle in the right side of Figure~\ref{fig:demonstration-most-study-map}). 
        These are road segments selected by the all the methods. 
    }
    \label{fig:demonstration-most-pickup-commonly}
\end{figure}

\noindent
{\bf Selected Road Segments.}
To save space, we only describe road segments selected by \codeName{MMD-CV-AGG}, which performed well in the previous experiments, on interval 1 (4:00 am - 5:23 am), interval 3 (6:46 am - 8:09 am) and interval 5 (9:32 am - 10:55 am) in Figure~\ref{fig:demonstration-most-pickup-by-mmd-cv-overview}. 
The purpose is to understand how the discrepancies between the two scenarios evolve over time.  
We can make the following observations:
\begin{itemize}
    \item {\bf Interval 1 (4:00 am - 5:23 am).} The {\tt A8} highway, indicated by the thick curve north of Monaco, is mainly selected. 
    (Other road segments are also selected, but they are scattered.)
    The vehicles that travel on the {\tt A8} highway in the original scenario cannot pass there due to the road blocking in the modified scenario, and this discrepancy appears to be captured by the selected road segments. 

    \item {\bf Interval 3 (6:46 am - 8:09 am).}  In addition to the {\tt A8} highway, many other road segments are selected. 
    Particularly, the road running to the south direction from the west side of {\tt A8}, the roads around the east side of {\tt A8}, and the roads along the {\tt A8}, are selected. 
    They correspond to the roads indicated by the light blue, orange and purple arrows in Figure~\ref{fig:demonstration-most-study-map}. 
    Moreover, more roads are selected in Monaco city than interval 1. 
    
    The traffic on these roads would have increased because the vehicles that could not pass {\tt A8} changed their routes to bypass {\tt A8} in the modified scenario.  
    Indeed, Figure~\ref{fig:demonstration-most-pickup-commonly} (left) shows the increase of the traffic on a road segment along the {\tt A8} highway (near the left purple arrow in Figure~\ref{fig:demonstration-most-study-map}) in the modified scenario.
    Figure~\ref{fig:demonstration-most-pickup-commonly} (right) shows a similar increase on a road segment on the east side of {\tt A8} (near the right blue circle in Figure~\ref{fig:demonstration-most-study-map}). 
    These discrepancies are captured by the selected road segments. 

    \item {\bf Interval 5 (9:32 am - 10:55 am).}  
    Here, while the {\tt A8} highway is still selected, the north-south road at the west side of {\tt A8} (the left light-blue arrows in Figure~\ref{fig:demonstration-most-study-map}), which was selected on interval 3, is not selected. 
    Similarly, fewer road segments are selected along the route bypassing {\tt A8} (the purple arrows in Figure~\ref{fig:demonstration-most-study-map}). 
This would be because the vehicles that increased traffic on interval 3 moved to the east. Indeed, many road segments on the east side are selected.
 \end{itemize}

\begin{figure}[t]
    \centering
    \includegraphics[scale=0.4]{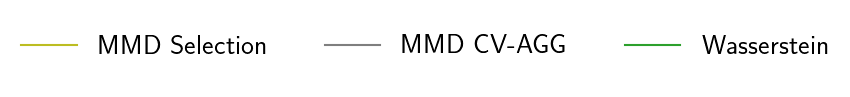} \hfill
    \\ 
    \includegraphics[width=.4\textwidth]{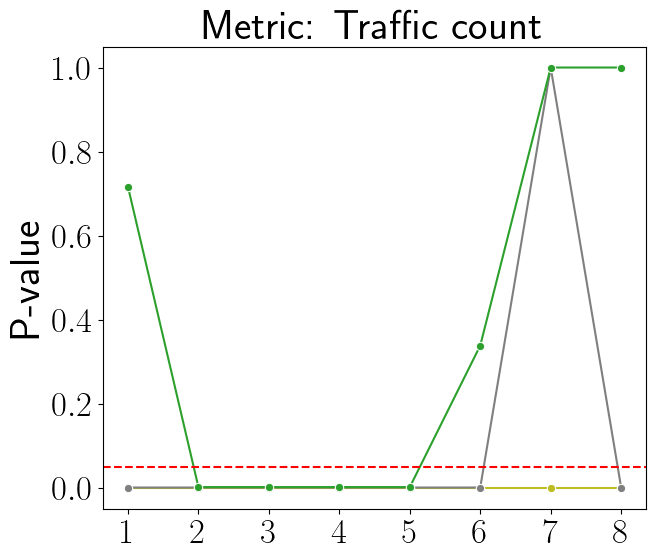} \ 
    \includegraphics[width=.4\textwidth]{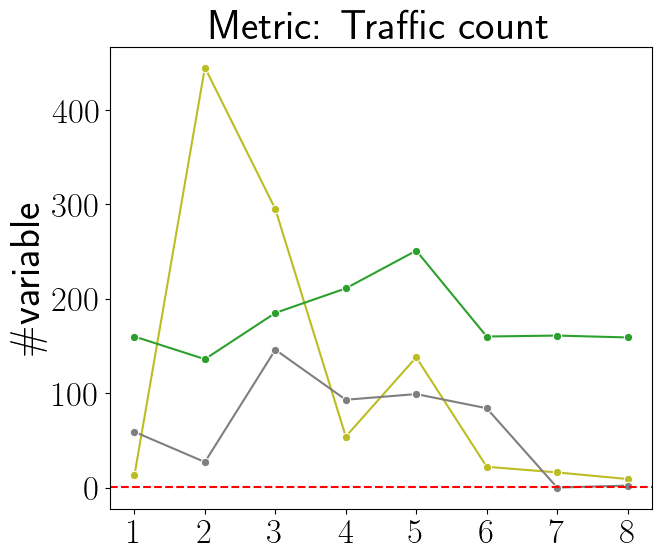} \hfill
    \caption{
        P-values~(left) and the number of selected variables~(right) of the three variable selection methods in Section~\ref{sec:demonstration-most}.
        The red dashed line represents the value $0.05$ in the left figure and zero  in the right figure.
    }
    \label{fig:demonstration-most-simple-aggregation-and-pval}
\end{figure}

\noindent
{\bf Permutation Tests.} 
Figure~\ref{fig:demonstration-most-simple-aggregation-and-pval}~(left) shows p-values on each subinterval obtained with the three variable selection methods.
From interval 2 to interval 5, all the methods resulted in low p-values, suggesting that the original and modified scenarios differ significantly on these subintervals.  
On the rest of intervals, {\tt Wasserstein} yielded high p-values, the reason of which may be that {\tt Wasserstein} tends to select more redundant variables (thus leading to leading low precision), as the previous experiments suggest. 
On interval 7, {\tt MMD-CV-AGG} selected no variable (as shown in Figure~\ref{fig:demonstration-most-simple-aggregation-and-pval} (right)), and it returned p-value $1.0$ (by design of the algorithm; see Appendix~\ref{sec:appendix-mmd-opt-when-null-hypothesis}).

\section{Conclusion}

We proposed a variable selection framework to compare two multivariate time-series data.  
It splits the entire time interval into subintervals and then applies a two-sample variable selection method on each subinterval. 
In this way, one can identify subintervals where the two time-series differ significantly and select variables responsible for their differences.

The proposed framework's strength is that it does not require multiple trajectories for each of the two-time series data, but only requires one realisation from each time series. 
This is particularly useful in applications where obtaining multiple trajectories is challenging, such as when one (or both) of the time-series data is from a computationally expensive simulator. 
We showcased the proposed framework's practical applications in fluid simulations, where we compared outputs from a simulator and its DNN emulator, and in traffic simulations, where we compared an original scenario with a modified one. 

One limitation of the framework is that if the number of subintervals is too small and hence each subinterval becomes long, then it may become difficult to detect subtle differences between two time-series, as we demonstrated in the synthetic data experiments. 
On the other hand, if the number of subintervals is too large so that each subinterval is short, then the number of samples on each interval becomes small, and thus, detecting discrepancies between the two data may also become challenging. 
Thus, to balance this trade-off, one needs to choose the number of subintervals (or the length of each subinterval) appropriately. 
Future work should theoretically investigate the framework, focusing particularly on this trade-off and the optimal choice of subintervals.

\paragraph{Acknowledgments}
The authors thank Paolo Papotti for a helpful discussion. 
A substantial part of the work was done during the first author's internship at Huawei Munich Research Center.

\bibliographystyle{plainnat}
\bibliography{main}

\appendix

\section{Details of MMD-based Two-Sample Variable Selection}
\label{sec:app-MMD}

We explain the details of the algorithmic settings of the MMD-based two-sample variable selection methods of \citet{mitsuzawa-2023} used in our experiments. 

\subsection{Optimisation of ARD Weights}

\label{sec:experiment-mmd-configuration}

The optimisation problem~\eqref{eq:mmd-optimisation-problem} for the ARD weights $a_1, \dots, a_n$ is numerically solved by using the {\tt Adam} optimiser \citep{Kingma-adam} implemented in {\tt PyTorch}~\citep{paszke2019pytorch}. \\

\noindent
{\bf Learning Rate Scheduler.}
We use the learning rate scheduler {\tt ReduceLROnPlateau}\footnote{\url{https://pytorch.org/docs/stable/generated/torch.optim.lr_scheduler.ReduceLROnPlateau.html}} to set the learning rate of the {\tt Adam} optimiser.
The initial value of the learning rate is set to $0.01$.
The scheduler monitors the value of the objective function~\eqref{eq:mmd-optimisation-problem} for every 10 epochs, reduces the learning rate by a factor of $0.5$ if the objective value does not dramatically change during these 10 epochs, and continues this until the learning rate reaches $0.001$. \\

\noindent
{\bf Early Stopping Criteria.}
We use two early stopping criteria defined as follows:
\begin{enumerate}
    \item  {\bf Convergence-based stopper:}
    This stopping criterion starts monitoring the objective value~\eqref{eq:mmd-optimisation-problem} after the initial 200 epochs. 
    It stops the optimisation when the objective value does not change significantly for the past $100$ epochs. 
    More specifically, the optimisation is stopped if the minimum of the objective value in the past $100$ epochs divided by the maximum of the objective value in the past $100$ epochs is less than $0.001$.
    
    \item {\bf Variable-selection-based stopper:} 
After the initial 400 epochs, the stopper starts performing variable selection every 10 epochs (by the thresholding algorithm applied to the ARD weights).  
    It stops the optimisation if the selected variables have not changed for the past $100$ epochs. 
\end{enumerate}
If neither criterion is triggered, the optimisation stops after $9,999$ epochs.

From our experience, the second stopping criterion is necessary. 
This is because the objective value may keep decreasing even when some ARD weights are already significant and the other weights are almost zero.
In such a case, the significant ARD weights keep increasing and the objective value does not converge. 
The second criterion prevents such unnecessary optimisation. \\

\noindent
{\bf When the MMD Estimate is Negative.}
\label{sec:appendix-mmd-opt-when-null-hypothesis}
When the probability distributions $P$ and $Q$ generating the datasets ${\bf X}$ and ${\bf Y}$ are identical or very similar, the MMD estimate~\eqref{eq:mmd-unbiased-est} can become negative because it is an unbiased estimator. 
In this case, the quantity~$\ell(a_1, \threeDots, a_{\Dim})$ in the optimisation problem~\eqref{eq:mmd-optimisation-problem} becomes negative, and thus its logarithm, which defines the objective function, is ill-defined.

To fix this issue, when the MMD estimate is negative, we replace the $\log( \ell(a_1, \dots, a_D) )$ by $\ell(a_1, \dots, a_D)$ in the objective~\eqref{eq:mmd-optimisation-problem}. 
Once it becomes positive during optimisation, the original form of \eqref{eq:mmd-optimisation-problem} is used.
If $\ell(a_1,...,a_{\Dim})$ keeps being negative for 3000 epochs, we stop the optimisation and accept the null hypothesis $P = Q$.
In this case, we return the empty set as the selected variables and set the p-value as $1.0$.

\subsection{Automatic Regularisation Parameter Selection}

We describe how the regularisation parameter $\lambda$ in \eqref{eq:mmd-optimisation-problem} is chosen in \codeName{MMD-Selection} and \codeName{MMD-CV-AGG}.

\subsubsection{\codeName{MMD-CV-AGG}}

For the \codeName{MMD-CV-AGG} algorithm of \citet{mitsuzawa-2023}, an upper bound $\lambda_{\rm upper}$ and a lower bound $\lambda_{\rm lower}$ for $\lambda$ should be specified.\footnote{In \citet{mitsuzawa-2023}, 
$\lambda_{\rm lower}$ is set to $0.01$ by default.
Setting $\lambda$ as $\lambda_{\rm lower}$ for initialisation and defining a step size $s > 0$, which is a hyperparameter, the following iteration procedure is applied until the number of selected variables $\hat{S} \subset \{1, \dots, D\}$ becomes $1$: Increase $\lambda$ by adding $s$, optimise with $\lambda$ and perform variable selection to obtain $\hat{S}$.
The resulting $\lambda$ is the value of $\lambda_{\rm upper}$.
The lower bound $\lambda_{\rm lower}$ and the step size $s$ are the hyperparameters of this procedure, but appropriately setting them is not straightforward. 
In particular, if $s$ is too small, the number of iterations increases significantly, while if $s$ is too large, $\lambda_{\rm upper}$ may become unnecessarily large. 
Therefore, the present paper uses the following improved heuristic algorithm.}  
Here we explain how $\lambda_{\rm upper}$ 
 and $\lambda_{\rm lower}$ are selected in the present paper.

We set $\lambda_{\rm upper}$ and $\lambda_{\rm lower}$ by separately optimising them using {\tt Optuna}, a hyperparameter optimisation framework based on Bayesian optimisation and efficient sampling~\citep{akiba2019optuna}. 
We optimise $\lambda_{\rm upper}$ in the range $[0.01, 2.0]$ to minimise the number $|\hat{S}|$ of selected variables, while we optimise  $\lambda_{\rm lower}$ in the range $[10^{-6}, 0.01]$ to maximise $|\hat{S}|$.
After $\lambda_{\rm upper}$ and  $\lambda_{\rm lower}$ are obtained, candidate values for $\lambda$ are calculated as $N_\lambda \in \mathbb{N}$ values equally spaced between $\lambda_{\rm lower}$ and  $\lambda_{\rm upper}$. 
(We set $N_\lambda = 10$ as a default value.)
All of these values are used in the aggregation algorithm of {\tt MMD-CV-AGG}. See \citet{mitsuzawa-2023} for further details.

\subsubsection{\codeName{MMD-Selection}} 
For \codeName{MMD-Selection}, the regularisation parameter $\lambda$ is optimised to find the best value. 
In \cite{mitsuzawa-2023}, $\lambda$ is selected from a specified set of candidate values, which may not be straightforward to specify. 
Therefore, in the present paper, $\lambda$ is optimised using {\tt Optuna}, as outlined in Algorithm~\ref{alg-model-model-selection}.
It seeks $\lambda$ that results in the ARD weights $a_\lambda$ with which (1) the MMD two-sample test has high power (as measured by $\ell_{\rm val}(a_\lambda)$), and (2) the selected variables $\hat{S}_\lambda$ based on $a_\lambda$ lead to a permutation test with a small p-value $p_\lambda$. 
See \cite{mitsuzawa-2023} for the definitions of notation and the mechanism of the algorithm.

We briefly explain the procedure of Algorithm~\ref{alg-model-model-selection}. 
Given datasets ${\bf X} = \{\bx_1, \dots, \bx_n\} \subset \mathbb{R}^D$ and ${\bf Y} = \{\by_1, \dots, \by_n\} \subset \mathbb{R}^D$ are divided into ``training'' sets ${\bf X}_{\rm train}$ and ${\bf Y}_{\rm train}$, which are used for optimising the ARD weights, and ``validation'' sets ${\bf X}_{\rm val}$ and ${\bf Y}_{\rm val}$, used for evaluating the test power proxy $\ell_{\rm val}(a_\lambda)$.
The algorithm requires the number $N_{\rm search}$ of iterations for {\tt Optuna}, which we set $N_{\rm search}=20$ in the experiments.

Line 1 initialises $\lambda^*$ and the objective value $\ell^*_{\rm Optuna}$ for {\tt Optuna}.
Line 4 numerically optimises the ARD weights $a_{\lambda}$ using a candidate $\lambda$ and the training sets.
Line 5 computes the test power proxy using the optimised ARD weights on the validation sets; see \citet[Equation (5)]{mitsuzawa-2023} for its exact form.
Line 7 computes p-value $p_\lambda$ by performing a permutation two-sample test on the validation sets using the selected variables $\hat{S}_{\lambda}$ from Line 6.
Line 8 updates the {\tt Optuna} objective as $\ell_{\rm val}(a_\lambda) (1 - p_\lambda)$. 
If it is higher than the current best value $\ell^*_{\rm Optuna}$, update it as the new best value (Lines 9-11).  

Lastly, the algorithm outputs the selected variables $\hat{S}_{\lambda^*}$ and the best regularisation parameter $\lambda^*$.

\begin{algorithm}[t]
    \caption{
        Data-driven Regularisation Parameter Selection for {\tt MMD-Selection}
        \label{alg-model-model-selection}
        \newline
        \Comment{
            \textbf{Input:} $N_{\rm search}$: the number of iterations for {\tt Optuna} ($N_{\rm search} = 20$ as default).
            $\Lambda \subset \mathbb{R}_+$: search space ($\Lambda =  [0.000001, 2.0]$ as default).
            $({\bf X}_{\rm train}, {\bf Y}_{\rm train})$: training data. 
            $({\bf X}_{\rm val}, {\bf Y}_{\rm val})$: validation data.
        }
        \newline
        \Comment{
            \textbf{Output:} $\hat{S}_{\lambda^*} \subset \{1, \dots, D\}$: selected variables with  the best regularisation parameter $\lambda^* \in \Lambda$.
        }
        \newline
    }
    \begin{algorithmic}[1]
        \State Initialise $\lambda^*$ using \verb|suggest_float()| and the {\tt Optuna} objective  $\ell^*_{\rm Optuna} = 0.0$.
        \While {$N_{\rm search}$ times}
            \State {\tt Optuna} suggests new $\lambda$.

            \State Obtain ARD weights $a_{\lambda} \in \mathbb{R}^D$ by  numerically solving \eqref{eq:mmd-optimisation-problem} using $({\bf X}_{\rm train}, {\bf Y}_{\rm train})$ and $\lambda$.
  
            \State Compute $\ell_{\rm val}(a_\lambda) > 0$ on $({\bf X}_{\rm val}, {\bf Y}_{\rm val})$.
         
            \State Select variables $\hat{S}_{\lambda} \subset \{1, \dots, D\}$ using $a_{\lambda}$.
  
            \State Compute p-value $0 \leq p_\lambda \leq 1$ by a permutation two-sample test on $({\bf X}_{\rm val}[:, \hat{S}_{\lambda} ], {\bf Y}_{\rm val}[:, \hat{S}_{\lambda} ])$.

            \State Update the {\tt Optuna} objective $\ell_{\rm Optuna} = (1 - p_\lambda) \ell_{\rm val}(a_\lambda)$.

            \If {$\ell_{\rm Optuna} > \ell^*_{\rm Optuna}$}
                \State Update $\ell^*_{\rm Optuna} = \ell_{\rm Optuna}$ and $\lambda^* = \lambda$.
            \EndIf
      \EndWhile
    \end{algorithmic}  
\end{algorithm}

\end{document}